\documentclass[10pt]{article}
\usepackage{bbm}
\usepackage[final]{graphics}
\usepackage{amsmath}
\usepackage{amsfonts,amsbsy}
\usepackage{amssymb}

\def\empile#1\over#2{\mathrel{\mathop{\kern 0pt#1}\limits_{#2}}}
\def\bs{\boldsymbol}
\def\wt#1{\widetilde{#1}}

\newcommand{\slv}{\raise.15ex\hbox{$/$}\kern-.53em\hbox{$v$}}
\newcommand{\slF}{\raise.15ex\hbox{$/$}\kern-.53em\hbox{$F$}}
\newcommand{\slL}{\raise.15ex\hbox{$/$}\kern-.53em\hbox{$L$}}
\newcommand{\slP}{\raise.15ex\hbox{$/$}\kern-.53em\hbox{$P$}}
\newcommand{\slp}{\raise.15ex\hbox{$/$}\kern-.53em\hbox{$p$}}
\newcommand{\slq}{\raise.15ex\hbox{$/$}\kern-.53em\hbox{$q$}}
\newcommand{\slR}{\raise.15ex\hbox{$/$}\kern-.53em\hbox{$R$}}
\newcommand{\slQ}{\raise.15ex\hbox{$/$}\kern-.53em\hbox{$Q$}}
\newcommand{\slK}{\raise.15ex\hbox{$/$}\kern-.53em\hbox{$K$}}
\newcommand{\slk}{\raise.15ex\hbox{$/$}\kern-.53em\hbox{$k$}}
\newcommand{\slD}{\raise.15ex\hbox{$/$}\kern-.73em\hbox{$D$}}
\newcommand{\slC}{\raise.15ex\hbox{$/$}\kern-.53em\hbox{$C$}}
\newcommand{\slA}{\raise.15ex\hbox{$/$}\kern-.53em\hbox{$A$}}
\newcommand{\slSigma}{\raise.15ex\hbox{$/$}\kern-.53em\hbox{$\Sigma$}}
\newcommand{\slpartial}{\raise.15ex\hbox{$/$}\kern-.53em\hbox{$\partial$}}
\newcommand{\slcalP}{\raise.15ex\hbox{$/$}\kern-.63em\hbox{$\cal P$}}

\def\p{{\boldsymbol p}}

\def\k{{\boldsymbol k}}

\def\x{{\boldsymbol x}}
\def\y{{\boldsymbol y}}

\def\v{{\boldsymbol v}}

\def\u{{\boldsymbol u}}

\newcommand{\as}{\alpha_\mathrm{s}}
\newcommand{\lqcd}{\Lambda_{_{\rm QCD}}}
\newcommand{\qs}{{Q_s}}
\newcommand{\opt}{\mathbbm{T}}

\catcode`\@=11

\newcount\@tempcntc
\def\@citex[#1]#2{\if@filesw\immediate\write\@auxout{\string\citation{#2}}\fi
  \@tempcnta\z@\@tempcntb\m@ne\def\@citea{}\@cite{%
        \@for\@citeb:=#2\do%
    {\@ifundefined{b@\@citeb}%
        {\@citeo\@tempcntb\m@ne\@citea%
                \def\@citea{,\penalty\@m\ }{\bf ?}\@warning%
                {Citation `\@citeb' on page \thepage \space undefined}}%
        {\setbox\z@\hbox{\global\@tempcntc0\csname b@\@citeb\endcsname\relax}
     \ifnum\@tempcntc=\z@ \@citeo\@tempcntb\m@ne%
       \@citea\def\@citea{,\penalty\@m}%
       \hbox{\csname b@\@citeb\endcsname}%
     \else%
      \advance\@tempcntb\@ne%
      \ifnum\@tempcntb=\@tempcntc%
      \else\advance\@tempcntb\m@ne\@citeo%
      \@tempcnta\@tempcntc\@tempcntb\@tempcntc\fi\fi}}\@citeo}{#1}}%

\def\@citeo{\ifnum\@tempcnta>\@tempcntb\else\@citea
  \def\@citea{,\penalty\@m}%
  \ifnum\@tempcnta=\@tempcntb\the\@tempcnta\else
   {\advance\@tempcnta\@ne\ifnum\@tempcnta=\@tempcntb \else
\def\@citea{--}\fi
    \advance\@tempcnta\m@ne\the\@tempcnta\@citea\the\@tempcntb}\fi\fi}

\catcode`\@=12

\begin{document}

\title{\bf High energy factorization\\ in nucleus-nucleus collisions}

\date{}

\author{Fran\c cois Gelis$^{(1)}$, Tuomas Lappi$^{(2)}$, Raju Venugopalan$^{(3)}$}
\maketitle
\begin{center}
\begin{enumerate}
\item Theory Division, PH-TH, Case C01600, CERN,\\
 CH-1211, Geneva 23, Switzerland
\item Institut de Physique Th\'eorique (URA 2306 du CNRS)\\
  CEA/DSM/Saclay, B\^at. 774\\
  91191, Gif-sur-Yvette Cedex, France
\item  Physics Department, Brookhaven National Laboratory\\
  Upton, NY-11973, USA
\end{enumerate}
\end{center}

\maketitle

\maketitle

\begin{abstract}
  We derive a high energy factorization theorem for inclusive gluon
  production in A+A collisions. Our factorized formula resums i) all
  order leading logarithms $(g^2\,\ln(1/x_{1,2}))^n$ of the incoming
  partons momentum fractions, and ii) all contributions $(g
  \rho_{1,2})^n$ that are enhanced when the color charge densities in
  the two nuclei are of order of the inverse coupling--
  $\rho_{1,2}\sim g^{-1}$. The resummed inclusive gluon spectrum can
  be expressed as a convolution of gauge invariant distributions
  $W[\rho_{1,2}]$ from each of the nuclei with the leading order gluon
  number operator.  These distributions are shown to satisfy the
  JIMWLK equation describing the evolution of nuclear wavefunctions
  with rapidity. As a by-product, we demonstrate that the JIMWLK
  Hamiltonian can be derived entirely in terms of retarded light cone
  Green's functions without any ambiguities in their pole
  prescriptions.  We comment on the implications of our results for
  understanding the Glasma produced at early times in A+A collisions
  at collider energies.

\end{abstract}
\begin{flushright}
Preprint IPhT-T08/068, CERN-PH-TH-2008-074.
\end{flushright}

\section{Introduction}

Collinear factorization theorems~\cite{Muell10} that isolate long
distance non-perturbative parton distribution functions from
perturbatively calculable short distance matrix elements are central
to the predictive power and success of QCD. These theorems can be
applied to compute inclusive cross-sections of the form
$A+B\longrightarrow I(M) + X$, where $I(M)$ is a set of heavy
particles or jets with invariant mass $M$ and $X$ corresponds to the
sum over all possible states (including soft and collinear hadrons)
that can accompany the object $I(M)$.  This cross-section, for center
of mass energy $\sqrt{s}$, can be expressed
as~\cite{ColliSS1,ColliSS2,ColliSS3,ColliSS4,ColliSS5}
\begin{eqnarray}
&&\sigma_{_{\rm{AB}}} = \sum_{ab} \int dx_a dx_b \;
f_{a/A}(x_a,\mu^2) f_{b/B}(x_b,\mu^2)\nonumber\\
&&\qquad\qquad\qquad\qquad\times\;
 {\hat \sigma}_{ab}
\left(\frac{M^2}{x_a x_b s},\frac{M}{\mu},\as(\mu)\right)\;
\left(1+ {\cal O}\left(\frac{1}{M^n}\right)\right) \; .
\label{eq:coll-fact}
\end{eqnarray}
In this equation, $f_{a (b)/A (B)}(x_{a(b)},\mu^2)$ are the
non-perturbative ``leading twist'' parton distribution functions which
gives the distribution of a parton $a(b)$ in the hadron $A(B)$, as a
function of the longitudinal momentum fraction $x_{a(b)}$ evolved up
to the factorization scale $\mu^2$, while the hard scattering matrix
element ${\hat \sigma_{ab}}$ can be computed systematically in a
perturbative expansion in powers of $\as= g^2/4\pi$, where $g$ is the
QCD coupling constant. Higher twist contributions to this formula are
suppressed by powers $n$ of the hard scale $M$. This factorization
formula is valid in the Bjorken limit when $M^2 \sim s \gg \lqcd^2$
(where $\lqcd \sim 200$~MeV is the intrinsic QCD scale).

Our interest here is instead in a different regime of high energy
scattering where, for fixed invariant mass $M\gg \lqcd$, one takes
$\sqrt{s}\rightarrow \infty$ and thus $x_{a,b}\rightarrow 0$.  We
shall call this the Regge--Gribov limit of QCD. An important insight
is that in this limit the field strengths squared can become very
large (${\cal O}(\frac{1}{\as})$) corresponding to the saturation of
gluon densities~\cite{GriboLR1,MuellQ1}. The onset of saturation is
characterized by a saturation scale $\qs(x)$, which opens a kinematic
window $M^2 \sim \qs^2 \gg \lqcd^2$ accessible at very high energies.
The physics of the Regge--Gribov regime is quite different from that
of the Bjorken limit discussed previously. The typical momenta of
partons are $\sim \qs \gg \lqcd$ and higher twist contributions are
not suppressed. These considerations are especially relevant for the
scattering of large nuclei because the large transverse density of
partons in the nuclear wavefunctions (proportional to the nuclear
radius $\sim A^{1/3}$) provides a natural enhancement of the
saturation scale, $Q_s^2(x,A) \propto A^{1/3}$. Our goal is to derive a
formula similar to eq.~(\ref{eq:coll-fact}) for inclusive gluon
production in the Regge--Gribov limit.

The dynamics of large parton phase space densities in the
Regge--Gribov limit can be described in the Color Glass Condensate
(CGC) effective field theory where small $x$ partons in hadrons and
nuclei are described by a classical field, while the large $x$ partons
act as color sources for the classical
field~\cite{McLerV1,McLerV2,McLerV3}.  The lack of dependence of
physical observables on the (arbitrary) separation between large $x$
color sources and small $x$ dynamical fields is exploited to derive a
renormalization group (RG) equation, known as the JIMWLK equation
\cite{JalilKMW1,JalilKLW1,JalilKLW2,JalilKLW3,JalilKLW4,IancuLM1,IancuLM2,FerreILM1}.
This equation is a functional RG equation describing the change in the
statistical distribution of color sources $W_{_Y}[\rho]$ with rapidity
$Y$ (= $\ln(1/x)$). It can be expressed as
\begin{equation}
{\partial W_{_Y}[\rho]\over \partial Y} = {\cal H}\; W_{_Y}[\rho] \; ,
\label{eq:H-JIMWLK1}
\end{equation}
where ${\cal H}$ is the JIMWLK Hamiltonian\footnote{The explicit form
of this Hamiltonian will be given later in the text.}. For a physical
observable defined by an average over all the source configurations,
\begin{equation}
\langle {\cal O}\rangle_{_Y} \equiv \int [D\rho]\; W_{_Y}[\rho]\;
 {\cal O}[\rho] \; ,
\label{eq:CGC-avg}
\end{equation}
one obtains 
\begin{equation}
{\partial \langle {\cal O}\rangle_{_Y}\over \partial Y} = \langle {\cal H}\;
 {\cal O}\rangle_{_Y} \; .
\label{eq:CGC-hier}
\end{equation}
We have used here eq.~(\ref{eq:H-JIMWLK1}) and integrated by parts
(using the hermiticity of ${\cal H}$). The structure of ${\cal H}$ is
such that $\langle {\cal H}{\cal O}\rangle_{_Y}$ is an object distinct
from $\langle {\cal O}\rangle_{_Y}$, so that one obtains in principle
an infinite hierarchy of evolution equations for operators expectation
values $\langle{\cal O}\rangle_{_Y}$~\cite{Balit1}. In the large $N_c$
and large $A$ mean-field limit, this hierarchy simplifies
greatly. When ${\cal O}$ is the ``dipole'' operator, corresponding to
the forward scattering amplitude in deep inelastic scattering, the
resulting closed evolution equation is known as the Balitsky-Kovchegov
(BK) equation~\cite{Balit2,Kovch2}.

In refs.~\cite{GelisV1,GelisV2,GelisV4}, we developed a formalism to
compute observables related to multiparticle production in field
theories with strong time dependent sources. This formalism is
naturally applicable to the CGC description of high energy
scattering\footnote{Although the color sources of each nucleus are
independent of the corresponding light-cone time, their sum
constitutes a time-dependent current.} albeit, for simplicity, we
considered only a scalar $\phi^3$ field theory.  (The corresponding
QCD framework was briefly considered in ref.~\cite{GelisLV2}.) In
these papers, the formalism for multiparticle production was developed
for a fixed distribution of sources, with the assumption that the
final results could be averaged over, as in eq.~(\ref{eq:CGC-avg}),
with unspecified distributions of sources $W_{_{Y_1}}[\rho_1]$ and
$W_{_{Y_2}}[\rho_2]$ (one for each of the projectiles). However, we
did not discuss in these papers the validity of such a factorization
formula.

In the formalism of  refs.~\cite{GelisV1,GelisV2,GelisV4}, one can formally arrange the perturbative expansion of an observable
like the single inclusive gluon spectrum as 
\begin{equation}
{\cal O}\left[\rho_1,\rho_2\right]
=
\frac{1}{g^2}\Big[c_0+c_1 g^2+c_2 g^4+\cdots\Big]\; ,
\label{eq:expansion}
\end{equation}
where each term corresponds to a different loop order. Each of the
coefficients $c_n$ is itself an infinite series of terms involving
arbitrary orders in $(g\rho_{1,2})^p$. We call ``Leading Order'' the
contribution that comes from the first coefficient $c_0$~:
\begin{equation}
{\cal O}_{_{\rm LO}}[\rho_1,\rho_2]\equiv\frac{c_0}{g^2}\; .
\end{equation}
In the case of the single gluon spectrum, the first term $c_0/g^2$ has
been studied extensively. In \cite{GelisV2} we developed
tools to calculate the next term $c_1$. Following this
terminology, we denote
\begin{equation}
{\cal O}_{_{\rm NLO}}[\rho_1,\rho_2]\equiv c_1\quad,\qquad
{\cal O}_{_{\rm NNLO}}[\rho_1,\rho_2]\equiv c_2\,g^2\; ,\cdots
\end{equation}
However, this strict loop expansion ignores the fact that large
logarithms of the momentum fractions $x_{1,2}$ can appear in the
higher order coefficients $c_{1,2,\cdots}$ when $\sqrt{s}$ is very
large.  The term $c_n$ can have up to $n$ powers of such logarithms,
and a more precise representation of these coefficients is 
\begin{equation}
c_n=\sum_{i=0}^n d_{ni}\,\ln^i\left(\frac{1}{x_{1,2}}\right)\; .
\label{eq:dni}
\end{equation}
The ``Leading Log'' terms are defined as those terms that have as many
logarithms as their order in $g^2$,
\begin{equation}
{\cal O}_{_{\rm LLog}}[\rho_1,\rho_2]\equiv\frac{1}{g^2}
\sum_{n=0}^\infty
d_{nn}\,g^{2n}\,\ln^n\left(\frac{1}{x_{1,2}}\right)\; .
\label{eq:LLog}
\end{equation}
In this work, we will go significantly further than the Leading
Order result, and resum the complete series of Leading Log terms. We
will prove that, after averaging over the sources $\rho_{1,2}$, all
the Leading Log corrections are automatically resummed by the JIMWLK
evolution of the distribution of sources, and that the event averaged
Leading Log result is given by the factorized expression
\begin{equation}
\langle {\cal O}\rangle_{_{\rm LLog}}
= \int [D\rho_1] [D\rho_2]\;
W_{Y_{\rm beam}-Y}[\rho_1]\, W_{Y_{\rm beam}+Y}[\rho_2]\;
{\cal O}_{_{\rm LO}}\left[\rho_1,\rho_2\right] \; .
\label{eq:fact-formula}
\end{equation}
In this formula, $Y$ is the rapidity at which
the gluon is measured, and the subscripts $Y_{\rm beam}\pm Y$ indicate
the amount of rapidity evolution\footnote{In terms of the center of
mass energy $\sqrt{s}$ of the collision (for a nucleon-nucleon pair)
and the longitudinal momentum components $p^\pm$ of the measured
gluon, one has also -- at leading log -- $Y_{\rm beam}-
Y=\ln(\sqrt{s}/p^+)$ and $Y_{\rm beam}+ Y=\ln(\sqrt{s}/p^-)$.} of the
source distributions of the two projectiles, starting in their
respective fragmentation regions.

The expressions $W_{Y_{\rm beam}\pm Y}[\rho_{1,2}]$ in
eq.~(\ref{eq:fact-formula}) are gauge invariant functionals describing
the source distributions in each of the nuclei.  In analogy to the
parton distribution functions $f_{a (b)/A (B)}(x_{a(b)},\mu^2)$ we
 introduced previously, they contain non-perturbative information
on the distribution of sources at rapidities close to the beam
rapidities. Just as the latter evolve in $\mu^2$ with the
DGLAP~\cite{GriboL2,AltarP1,Doksh1} evolution equations, the former,
as suggested by eq.~\ref{eq:H-JIMWLK1}, obey the JIMWLK evolution
equation in rapidity which evolves them up to the rapidities $Y_{\rm
  beam}- Y$ and $Y_{\rm beam}+ Y$ from the nuclei $A_1$ and $A_2$
respectively.
As we will discuss in detail, the leading order inclusive gluon
spectrum, for given sources $\rho_{1,2}$, can be computed by solving
the classical Yang-Mills equations with simple retarded boundary
conditions. Eq.~(\ref{eq:fact-formula}) suggests that the result
resumming all the leading logarithms of the collision energy can be
obtained by averaging over this leading order result with the weight
functionals $W$ evolved from the beam rapidity to the rapidity $Y$ at
which the gluon is produced. 

In the Regge--Gribov limit, eq.~(\ref{eq:fact-formula}) is the analog
of the factorization formula eq.~(\ref{eq:coll-fact}) proved in the
Bjorken limit. While we will prove that eq.~(\ref{eq:fact-formula})
holds for leading logarithmic contributions at all orders in
perturbation theory, we have not attempted to show that it is valid
for sub-leading logarithms. There is currently an intense activity in
computing sub leading logarithmic contributions in the high parton
density
limit~\cite{Balit3,BalitC1,GardiKRW1,KovchW1,KovchW2,KovchW3,AlbacK1}
so an extension of our results beyond leading logs is feasible in future. 
There is another aspect of A+A collisions that we have not
discussed thus far. Our power counting does not account for the so
called ``secular divergences''~\cite{Golde1,GelisJV1,Berges:2004yj}. 
These are
contributions that diverge at least as powers of the time elapsed 
after the collision. Including these contributions will not alter our
factorization theorem; it does affect how ``observables'' defined at
finite times {\it after} the nuclear collisions are related to
quantities measured in A+A experiments. We will address this issue
briefly. A fuller treatment requires more work.

The paper is organized as follows. In section \ref{sec:nlo}, we derive
an important formula for the Next to Leading Order corrections to the
inclusive gluon spectrum. This formula will play a crucial role later,
in disentangling the initial state effects from the rest of the
collision process. In section \ref{sec:1nuc}, we will derive the
expressions stated in eqs.~(\ref{eq:H-JIMWLK1})--(\ref{eq:CGC-hier})
for JIMWLK evolution of a single nucleus. Albeit the result is well
known, our derivation is quite different from those existing in the
literature~\cite{JalilKMW1,JalilKLW1,JalilKLW2,JalilKLW3,JalilKLW4,IancuLM1,IancuLM2,FerreILM1,KovneM1,Weige1,Muell5,KovneW1,MuellSW1}.
We will obtain our result entirely in terms of retarded light-cone
Green's functions without any recourse to time-ordered propagators. We
will show that there are no ambiguities in specifying the pole
prescriptions in this approach. More importantly, our derivation
allows us to straightforwardly extend our treatment of the JIMWLK
equation to the case of the collision of two nuclei. This is discussed
separately in section \ref{sec:2nuc} where we show explicitly that
non-factorizable terms are suppressed and our key result, stated in
eq.~(\ref{eq:fact-formula}), is obtained.  In the following section,
we will relate our work to previous work in this direction and briefly
explore some of the connections between the different approaches. In
section \ref{sec:glasma}, we will discuss how one can relate our
result for the Glasma produced at early times in heavy ion
collisions~\cite{LappiM1,GelisV4} and its subsequent evolution into
the Quark Gluon Plasma. We conclude with a brief summary and
discussion of open issues. There are three appendices dealing with
properties of Green's functions in light cone gauge relevant to the
discussion in the main text of the paper.

\section{NLO corrections to inclusive observables}
\label{sec:nlo}
Before studying the logarithmic divergences that arise in loop
corrections to observables, let us derive a formula that expresses the
1-loop corrections to inclusive observables in terms of the action of
a certain operator acting on the same observable at leading order. As
we shall see, this formula -- albeit quite formal -- can be used to
separate the physics of the initial state from the collision itself.

\begin{figure}[htbp]
\begin{center}
\resizebox*{!}{1cm}{\includegraphics{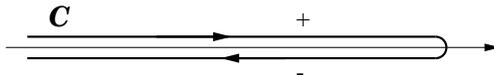}}
\end{center}
\caption{\label{fig:SK}The closed time path used in the
Schwinger-Keldysh formalism.}
\end{figure}

We have in mind an operator made of elementary color fields, which
probes multi-gluon correlations.  To be specific, for a given source
distribution, we shall consider the quantum expectation value
\begin{equation}
 {\cal O}(x,y)\equiv\big<A^i(x)A^j(y)\big>\; ,
\label{eq:op-def}
\end{equation}
in the limit where the time arguments of the two fields go to
$+\infty$. We chose this particular operator because we wish to study
the single gluon spectrum --the first moment of the multiplicity
distribution-- in the collision of two nuclei; it is obtained by
Fourier transforming this bilinear combination of fields. Note that
the two fields are not time-ordered. The expectation value of such a
product can be calculated in the Schwinger-Keldysh
formalism~\cite{Schwi1,Keldy1,BakshM1}, by considering that $A^i(x)$ lies
on the $-$ branch of the contour and $A^j(y)$ on the $+$ branch (A
representation of the Schwinger--Keldysh contour is shown in
fig.~\ref{fig:SK}.)
    
This section is organized as follows. We first recall the
expression of eq.~(\ref{eq:op-def}) at leading order in terms of
retarded solutions of the classical Yang--Mills equations. This result
is well known and has been derived in a number of different ways. We
will then discuss the next-to-leading order computation of this
quantity in the CGC framework. There are two sorts of NLO corrections;
these are the virtual corrections arising from one-loop corrections to
the classical fields and the real corrections which are obtained by
computing the ${\cal G}_{-+}$ propagator of a small fluctuation in
light-cone gauge. We will show that ${\cal O}_{_{\rm NLO}}$ can be
expressed as a linear operator with real and virtual pieces acting on
${\cal O}_{_{\rm LO}}$, plus an unimportant (as far as the resummation
of logs of $1/x_{1,2}$ is concerned) additional term.
 
\subsection{Leading order result}
We showed in \cite{GelisV1} that, at leading order, ${\cal O}$ is the
product of two classical solutions of the Yang-Mills equations, with
null retarded boundary conditions\footnote{The retarded nature of the
boundary conditions is intimately related to the inclusiveness of the
observable under consideration. For instance, if instead of the single
inclusive gluon spectrum, one wanted to calculate at leading order the
probability of producing a fixed number of gluons, one would have to 
solve the classical Yang-Mills equations with boundary
conditions both at $x^0=-\infty$ and at $x^0=+\infty$ (see
\cite{GelisV3}).},
\begin{equation}
{\cal O}_{_{\rm LO}}(x,y)={\cal A}^i(x){\cal A}^j(y)\; ,
\label{eq:O-LO}
\end{equation}
with
\begin{eqnarray}
\big[{\cal D}_\nu,{\cal F}^{\mu\nu}\big]&=&J^\nu\; ,
\nonumber\\
\lim_{x^0\to -\infty}{\cal A}^\mu(x)&=&0\; .
\label{eq:YM}
\end{eqnarray}
Here, ${\cal A}$ denotes the classical field, and $J^\mu$ is the color
current corresponding to a fixed configuration of the color sources.
The current is comprised of one or two sources depending on whether we consider
only one nucleus or the collision of two nuclei -- this distinction
is not important in this section. It is important to note that this current, which has
support only on the light-cone, must be covariantly conserved,
\begin{equation}
\big[{\cal D}_\mu,J^\mu\big]=0\; .
\end{equation}
This means that in general, there is a feed-back of the gauge field on
the current itself, unless one chooses a gauge condition such that the
gauge field does not couple to the non-zero components of the current
on the light-cone.

Although one can solve analytically the Yang-Mills equations with
these boundary conditions in the case of a single nucleus
\cite{JalilKMW1,Kovch1}, this is not possible in the case of two
nuclei, and one must resort to numerical methods to obtain results in
this case. Fortunately, as we shall see, the discussion of
factorization in the case of two nuclei does not require that we know
this solution analytically.

\begin{figure}[htbp]
\begin{center}
\resizebox*{!}{4cm}{\includegraphics{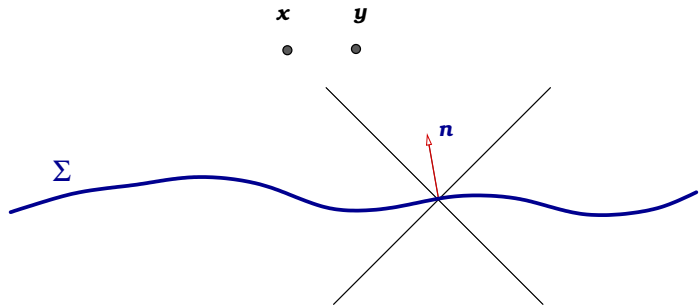}}
\end{center}
\caption{\label{fig:initialS}A locally space-like surface $\Sigma$
used to define the initial value of the color field.}
\end{figure}

Because the solution of the Yang-Mills equations we need is
defined with retarded boundary conditions, its value at the points $x$
and $y$ (where the observable is measured) is fully determined if we
know its value\footnote{Since the Yang-Mills equations contain second
derivatives with respect to time, one must also know the value of the
first time derivative of the field on this initial surface.} on an
initial surface $\Sigma$ --which is locally space-like\footnote{This
means that at every point $u\in\Sigma$, the vector $n^\mu$ normal to
$\Sigma$ at the point $u$ ($n^\mu dx_\mu=0$ for any displacement
$dx_\mu$ on $\Sigma$ around the point $u$) must be time-like. This
condition prevents a signal emitted at the point $u\in\Sigma$,
propagating at the speed of light, from encountering again the
surface $\Sigma$.}-- located below the points $x$ and $y$, as
illustrated in fig.~\ref{fig:initialS}.

Therefore, we will write
\begin{equation}
{\cal O}_{_{\rm LO}}(x,y)\equiv{\cal O}_{_{\rm LO}}[{\cal A}]\; ,
\end{equation}
which means that the observable is considered as a functional of the
value of the color field on the initial surface $\Sigma$. Note that we
use the same symbol for the color field and for its initial value on
$\Sigma$, although mathematically these objects depend on a different
number of variables and are therefore different functions.

\subsection{Next to leading order corrections}
A detailed discussion of the power counting for moments of the
inclusive multiplicity distribution can be found in
Ref.~\cite{GelisV1}. The leading order contributions to ${\cal
O}(x,y)$ involves only tree diagrams, which explains why it can be
obtained from classical solutions of the Yang-Mills equations. As
mentioned previously, this leading order contribution is of order
${\cal O}(\as^{-1})$ but includes all orders in $g\rho$. In
the rest of this section, we shall study the 1-loop corrections to
this quantity, that are of order ${\cal O}(1)$ in the coupling and to
all orders in $g\rho$.
 
The framework to compute these 1-loop corrections (hereafter called
``NLO'') to quantities such as eq.~(\ref{eq:op-def}) has been
developed for a scalar theory in ref.~\cite{GelisV2}. Much of this
analysis can be carried over to QCD. To avoid complications such as
ghost loops, we shall work in a gauge such as the light cone gauge
$A^+=0$. Following the discussion for the scalar case, we obtain at
NLO,
\begin{equation}
{\cal O}_{_{\rm NLO}}(x,y)=
{\cal A}^i(x)\,\beta^j(y)
+
\beta^i(x)\,{\cal A}^j(y)
+
{\cal G}^{ij}_{-+}(x,y)
\; .
\label{eq:O-NLO}
\end{equation}
In this equation, ${\cal G}^{ij}_{-+}(x,y)$ is the $-+$ component of
the small fluctuation Schwin\-ger-Keldysh propagator in the presence of
the classical background field ${\cal A}^i$ and the field $\beta^i$ is
the one loop correction to the classical field. It is obtained by
solving the small fluctuation equation of motion
\begin{equation}
\Big[
\square_x g^{\mu\nu}-\partial_x^\mu\partial_x^\nu
-
\frac{\partial^2 U({\cal A})}{\partial {\cal A}_\mu(x)\partial {\cal A}_\nu(x)}
\Big]\beta_\nu(x)
=
\frac{1}{2}
\frac{{\partial^3 U({\cal A})}}
{\partial {\cal A}_\mu(x)\partial {\cal A}^\nu(x) \partial {\cal A}^\rho(x)}
\,
{\cal G}^{\nu\rho}_{++}(x,x)\; ,
\label{eq:beta1}
\end{equation}
with null retarded boundary conditions~:
\begin{equation}
\lim_{x^0\to -\infty}\beta^\mu(x)=0\; .
\end{equation}
Here $U({\cal A})$ is the potential term in the Yang-Mills
Lagrangean\footnote{Unless one chooses a non-linear gauge condition,
  $U({\cal A})$ is made of the usual 3-gluon and 4-gluon couplings.},
obtained by writing
\begin{equation}
{\cal L} = {\cal L}_{\rm quad} - U({\cal A}) \, ,
\label{eq:QCD-Lag}
\end{equation}
where ${\cal L}_{\rm quad}$ is defined in eq.~(\ref{eq:L-quad}) of
appendix A. We refer the reader to appendix A for more details.  The
source term in this small fluctuation equation includes the closed
loop formed by the Schwinger-Keldysh propagator ${\cal G}_{++}(x,x)$
to be defined shortly, the third derivative corresponds to the 3-gluon
vertex in the presence of a background field and $1/2$ is a symmetry
factor.

Following \cite{GelisV2}, we can write the propagator ${\cal
G}^{ij}_{-+}(x,y)$ in eq.~(\ref{eq:O-NLO}) as a bilinear combination
of small fluctuations of the gauge field {\sl whose initial conditions are
plane waves},
\begin{equation}
{\cal G}^{ij,bc}_{-+}(x,y)
=
\sum_{\lambda,a}
\int\frac{d^3\k}{(2\pi)^3 2E_\k}\;
a^{ib}_{-\k \lambda a}(x)a^{jc}_{+\k \lambda a}(y)\; ,
\label{eq:G-+}
\end{equation}
where 
\begin{eqnarray}
&&
\Big[
\square_x g^{\mu\nu}-\partial_x^\mu\partial_x^\nu
-
\frac{\partial^2 U({\cal A})}{\partial {\cal A}_\mu(x)\partial {\cal A}_\nu(x)}
\Big]a_{\pm\k \lambda a,\nu}(x)
=
0
\; ,\label{eq:small-fluct-eqn}
\\
&&
\lim_{x^0\to-\infty}a^\mu_{\pm\k \lambda a}(x)=\epsilon^\mu_\lambda(\k)\,T^a\,e^{\pm ik\cdot x}\; .
\nonumber
\end{eqnarray}
The sum over $\lambda$ is over the two physical polarizations for the
initial plane wave and the index $a$ represents the initial color
carried by the small fluctuation field. In eq.~(\ref{eq:G-+}), our
notation is such that the lower color index ($a$) represents the
initial color of the fluctuation, while the upper color index ($b$ or
$c$) refer to its color after it has evolved on top of the classical
background field\footnote{For future reference, note that quantities
with only the lower color index are matrices in the adjoint
representation of $SU(N)$ defined by
\begin{equation}
a^\mu_{\pm\k \lambda a}(x)
\equiv
a^{\mu b}_{\pm\k \lambda a}(x)\,T^b\; .
\end{equation}
}.  It is important to stress that this decomposition of ${\cal
G}_{-+}^{ij}$ is valid only if one uses small fluctuations that are
plane waves in the remote past. Using other solutions of the small
fluctuation equation of motion (\ref{eq:small-fluct-eqn}) would lead
to a propagator that obeys incorrect boundary conditions.

The $++$ propagator at equal points can be written in a similar
fashion as\footnote{When the two end-points are separated by a
time-like interval, there can be an additional term contributing to
this propagator -- see \cite{GelisV2} for more general formulas.}
\begin{equation}
{\cal G}^{ij,bc}_{++}(x,x)
=
\sum_{\lambda,a}
\int\frac{d^3\k}{(2\pi)^3 2E_\k}\;
a^{ib}_{-\k \lambda a}(x)a^{jc}_{+\k \lambda a}(x)\; .
\label{eq:G++}
\end{equation}

We note that in a generic gauge, covariant current conservation may
require the incoming field fluctuation to induce a color precession of
the classical current $J^\mu$. This modification of the current will
in turn induce an additional contribution to the field
fluctuation. Our strategy~\cite{AyalaJMV1,AyalaJMV2,GelisM1} to avoid
this complication will be to perform all intermediate calculations in
a gauge where this phenomenon does not happen. For instance, on the
line $x^-=0$ where the color charges moving in the $+z$ direction
live, on should use a gauge in which ${\cal A}^-=0$. Indeed, because
the color current only has a $+$ component, covariant conservation is
trivial in this gauge.  A gauge rotation of the final result is then
performed to return to the light-cone gauge of interest. All effects
due to current conservation are then taken care of by this final gauge
transformation.

\subsection{Rearrangement of the NLO corrections - I}
\label{sec:massageI}
In this subsection, we will express the small fluctuation propagator
${\cal G}^{ij}_{-+}(x,y)$ as the action of a differential operator on
the classical fields ${\cal A}^i(x)$ and ${\cal A}^j(y)$. This
operator contains functional derivatives with respect to the initial
value of the color field on $\Sigma$. In the following subsection, we
will repeat the exercise for the one loop correction to the classical
field $\beta^\mu(x)$ and write it in terms of a similar operator
acting on the classical field ${\cal A}^\mu(x)$. These identities,
besides providing a transparent derivation of the JIMWLK equation for
a single nucleus, will be especially powerful in our treatment of
nucleus-nucleus collisions.

Let us begin from the Green's formula for the classical field ${\cal
  A}^\mu$,
\begin{eqnarray}
{\cal A}^\mu(x)= \int\limits_{\Sigma^+} d^4y\;
D^{\mu\rho}_{0,{}_R}(x,y)\, \frac{\partial U({\cal A})}{\partial {\cal
A}^\rho(y)} +{\cal B}^\mu_0[{\cal A}](x)\; ,
\label{eq:class-green}
\end{eqnarray}
where $D^{\mu\rho}_{0,{}_R}(x,y)$ is the free retarded propagator
(discussed in appendix A in the case of the light-cone gauge) and
${\cal B}^\mu_0[{\cal A}](x)$ is the boundary term that contains the
initial value of the classical field on $\Sigma$. (Boundary terms for
the classical and small fluctuation fields in light-cone gauge are
discussed in detail in appendix B.)  $\Sigma^+$ denotes the region of
space-time above the surface $\Sigma$. Now, consider an operator
$\bs{\cal T}$ (to be defined explicitly later) that acts on
the initial value of the fields on the surface $\Sigma$, and assume
that this operator is linear, which implies
\begin{equation}
\bs{\cal T} \frac{\partial U({\cal A})}{\partial {\cal A}^\rho(y)} 
= \frac{\partial^2 U({\cal A})}{\partial {\cal A}^\rho(y)\partial {\cal A}^\nu(y)} 
\bs{\cal T} {\cal A}^\nu(y)\; .
\end{equation}
Now apply this operator $\bs{\cal T}$ to both sides of
eq.~(\ref{eq:class-green}), we get
\begin{eqnarray}
\bs{\cal T} {\cal A}^\mu(x)
= \int\limits_{\Sigma^+} d^4y\;
D^{\mu\rho}_{0,{}_R}(x,y)\, 
\frac{\partial^2 U({\cal A})}{\partial {\cal A}^\rho(y)\partial {\cal A}^\nu(y)} 
\bs{\cal T} {\cal A}^\nu(y)
+\bs{\cal T}{\cal B}^\mu_0[{\cal A}](x)\; .
\end{eqnarray}
By comparing this equation with the Green's formula for a small
fluctuation $a^\mu$ (see appendix B),
\begin{eqnarray}
a^\mu(x)=
\int\limits_{\Sigma^+}d^4y\;D^{\mu\rho}_{0,{}_R}(x,y)
\frac{\partial^2 U({\cal A})}{\partial {\cal A}^\rho(y)\partial A^\nu(y)}
\;
a^\nu(y)+{\cal B}^\mu_0[a](x)\; ,
\label{eq:tmp4-green}
\end{eqnarray}
we see that we can identify $a^\mu(x)=\bs{\cal T}\,{\cal A}^\mu(x)$
provided that we have
\begin{equation}
{\cal B}^\mu_0[a](x)=\bs{\cal T}\,{\cal B}^\mu_0[{\cal A}](x)\; .
\end{equation}
Because ${\cal B}_0$ is a {\sl linear} functional of the initial value
of the color fields on the surface $\Sigma$, it is easy to see that
the operator $\bs{\cal T}$ that fulfils this goal is
\begin{equation}
\bs{\cal T}
\equiv
\int\limits_{\Sigma}d^3\vec\u\; \big[a\cdot
\opt_\u\big]\; ,
\end{equation}
where $\opt_\u$ is the generator of translations of the initial
fields\footnote{For now, it is sufficient to think of this operator as
  an operator which is linear in first derivatives with respect to the
  color field on $\Sigma$.}  at the point $\u\in\Sigma$. We denote by
$d^3\vec\u$ the measure on the surface $\Sigma$ (for instance, if
$\Sigma$ is a surface defined by $x^-={\rm const}$, this measure reads
$d^3\vec\u=du^+ d^2\u_\perp$.) The detailed expression of this
operator can be obtained by writing explicitly the Green's formula for
the retarded propagation of color fields above the surface $\Sigma$,
and it usually depends both on the choice of the surface and on the
choice of the gauge condition. An explicit expression of this operator
will be given in the next section when the initial surface $\Sigma$ is
parallel to the light-cone ($u^-={\rm const}$) and when the fields are
in the light-cone gauge ${\cal A}^+=0$.  Therefore, we have
established the following identity,
\begin{equation}
a^\mu(x)=\int\limits_{\Sigma}d^3\vec\u\; \big[a\cdot
\opt_\u\big]\;{\cal A}^\mu(x)\; .
\label{eq:aA}
\end{equation}
Eq.~(\ref{eq:aA}) provides a formal expression of a fluctuation at
point $x$ in terms of its value on some initial surface $\Sigma$ (in
the right hand side of eq.~(\ref{eq:aA}), only the value of the
fluctuation $a^\mu$ on $\Sigma$ appears). This formula is especially
useful in situations where we can calculate analytically the initial
value of the fluctuation on $\Sigma$, but were we do not know
analytically the classical background field ${\cal A}$ above this
surface. 

The single nucleus case is a bit academic in this respect
because one can analytically compute the background gauge field and
the fluctuation at any point in space-time.  Rather, eq.~(\ref{eq:aA}) will
prove especially powerful for nuclear collisions because in that case one does
not have an analytic expression for the classical background field
after the collision.

Armed with eq.~(\ref{eq:aA}), it is straightforward to write
the third term of the right hand side of eq.~(\ref{eq:O-NLO}) as
\begin{eqnarray}
{\cal G}^{ij,bc}_{-+}(x,y)
&=&
\sum_{\lambda,a}
\int\frac{d^3\k}{(2\pi)^3 2E_\k}\;
\int\limits_{\Sigma}
d^3\vec\u\, d^3\vec\v
\nonumber\\
&&\qquad\times
\Big[
\big[a_{-\k \lambda a}\cdot\opt_\u\big]
{\cal A}^{i b}(x)
\Big]
\Big[
\big[a_{+\k \lambda a}\cdot\opt_\v\big]
{\cal A}^{j c}(y)
\Big]\; .
\label{eq:green-NLO}
\end{eqnarray}
In this equation, the brackets limit the scope of the operators
${\mathbbm T_{\u,\v}}$.

\subsection{Rearrangement of the NLO corrections - II}
\label{sec:massageII}

The terms involving the 1-loop correction $\beta^\mu$ can also be
written in terms of the operator $\opt_\u$, but this is not as
straightforward as for ${\cal G}^{ij}_{-+}$. The first step is to
write down the formal Green's function solution of
eq.~(\ref{eq:beta1}). It is convenient to write it as
\begin{eqnarray}
&&
\!\!\!\!\!\!\!\!\!\!
\beta^\mu(x)
=
\underbrace{
\!\int\limits_{\Sigma^+}\!d^4y\;D^{\mu\nu}_{_R}(x,y)\;
\frac{1}{2}
\frac{{\partial^3 U({\cal A})}}
{\partial {\cal A}^\nu(y)\partial {\cal A}^\rho(y) \partial {\cal A}^\sigma(y)}
\,
{\cal G}^{\rho\sigma}_{++}(y,y)
}
+
\underbrace{\vphantom{\int\limits_{\Sigma}}
{\cal B}^\mu[\beta](x)
}
\; ,
\nonumber\\
&&\hskip 43mm\beta^\mu_1(x)
\hskip 40mm\beta^\mu_2(x)
\nonumber\\
&&
\label{eq:beta1+2}
\end{eqnarray}
where ${\cal B}^\mu[A](x)$ is identical to ${\cal B}^\mu_0[A](x)$
except that all occurrences of the bare propagator
$D_{0,_R}^{\mu\nu}$ in the latter  are replaced in the former by the dressed
propagator in the background field ${\cal A}^\mu$. This dressed
propagator, denoted $D_{_R}^{\mu\nu}$, satisfies the equation
\begin{equation}
\Big[
\square_x g^{\mu\nu}-\partial_x^\mu\partial_x^\nu
-
\frac{\partial^2 U({\cal A})}{\partial {\cal A}_\mu(x)\partial {\cal A}_\nu(x)}
\Big]D_{R,\mu}^\rho(x,y) = g^{\rho\nu} \delta(x-y) \; ,
\label{eq:green-beta}
\end{equation}
plus a retarded boundary condition such that it vanishes if $x^0<y^0$.

The second term on the right hand side of eq.~(\ref{eq:beta1+2}) is
the value $\beta$ would have if one turns off the source term
(proportional to ${\cal G}_{++}$) in the domain $\Sigma^+$ above the
initial surface. It is therefore given by a formula identical to
eq.~(\ref{eq:aA}),
\begin{equation}
\beta^\mu_2(x)
=
\int\limits_{\Sigma}d^3\vec\u\;
\big[\beta\cdot\opt_\u\big]\;{\cal A}^\mu(x)\; .
\end{equation}
To calculate $\beta_1(x)$, let us first make explicit
the interactions with the background field by writing it as
\begin{eqnarray}
\beta^\mu_1(x)
&=&
\!\int\limits_{\Sigma^+}d^4y\;D^{\mu\nu}_{0,{}_R}(x,y)
\Big[
\frac{{\partial^2 U({\cal A})}}
{\partial {\cal A}^\nu(y)\partial {\cal A}^\rho(y)}
\beta^\rho_1(y)
\nonumber\\
&&\qquad\qquad
+
\frac{1}{2}
\frac{{\partial^3 U({\cal A})}}
{\partial {\cal A}^\nu(y)\partial {\cal A}^\rho(y) \partial {\cal A}^\sigma(y)}
\,
{\cal G}^{\rho\sigma}_{++}(y,y)\Big]\; .
\label{eq:green-dA2}
\end{eqnarray}
This expression is obtained by substituting the expression for the
dressed retarded propagator in terms of the free retarded propagator
in the definition of $\beta^\mu_1$.

Consider now the quantity
\begin{equation}
\zeta^\mu(x)
\equiv\frac{1}{2}
\sum_{\lambda,a}
\int\frac{d^3\k}{(2\pi)^3 2E_\k}
\int\limits_{\Sigma}
d^3\vec\u\, d^3\vec\v\;
\big[a_{-\k \lambda a}\cdot\opt_\u\big]
\big[a_{+\k \lambda a}\cdot\opt_\v\big]
{\cal A}^\mu(x)\; .
\end{equation}
We shall prove that $\beta_1^\mu$ and $\zeta^\mu$ are
identical. Using eq.~(\ref{eq:aA}), we can write
\begin{equation}
\zeta^\mu(x)
=\frac{1}{2}
\sum_{\lambda,a}
\int\frac{d^3\k}{(2\pi)^3 2E_\k}
\int\limits_{\Sigma}
d^3\vec\u\;
\big[a_{-\k \lambda a}\cdot\opt_\u\big]
a_{+\k \lambda a}^\mu(x)\; .
\end{equation}
Replace $a_{+\k \lambda a}^\mu(x)$ in this equation by the r.h.s
of eq.~(\ref{eq:tmp4-green}). Because the boundary term ${\cal
B}_0^\mu[a_{+\k \lambda a}](x)$ does not depend on the initial value
of the classical field ${\cal A}$, the action of $\big[a_{-\k \lambda
a}\cdot\opt_\u\big]$ on this term gives zero. We thus obtain
\begin{eqnarray}
\zeta^\mu(x)
&=&
\frac{1}{2}
\sum_{\lambda,a}
\int\frac{d^3\k}{(2\pi)^3 2E_\k}
\int\limits_{\Sigma}
d^3\vec\u\;
\int\limits_{\Sigma^+}d^4y\;D^{\mu\nu}_{0,{}_R}(x,y)
\nonumber\\
&&
\qquad\times
\Big\{
\frac{{\partial^2 U({\cal A})}}
{\partial {\cal A}^\nu(y)\partial {\cal A}^\rho(y)}
\big[a_{-\k \lambda a}\cdot\opt_\u\big]a_{+\k \lambda a}^\rho(y)
\nonumber\\
&&\qquad\quad
+
\frac{{\partial^3 U({\cal A})}}
{\partial {\cal A}^\nu(y)\partial {\cal A}^\rho(y)\partial{\cal A}^\sigma(y)}
\Big[\big[a_{-\k \lambda a}\cdot\opt_\u\big]
{\cal A}^\sigma(y)\Big]
a_{+\k \lambda a}^\rho(y)
\Big\}
\nonumber\\
&=&
\int\limits_{\Sigma^+}d^4y\;D^{\mu\nu}_{0,{}_R}(x,y)
\Big[
\frac{{\partial^2 U({\cal A})}}
{\partial {\cal A}^\nu(y)\partial {\cal A}^\rho(y)}\;
\zeta^\rho(y)
\nonumber\\
&&\qquad\qquad\qquad\qquad\quad
+
\frac{{\partial^3 U({\cal A})}}
{\partial {\cal A}^\nu(y)\partial {\cal A}^\rho(y)\partial{\cal A}^\sigma(y)}
{\cal G}^{\rho\sigma}_{++}(y,y)
\Big]
\; ,
\end{eqnarray}
which is identical to eq.~(\ref{eq:green-dA2}). We therefore obtain
$\beta_1^\mu(x)=\zeta^\mu(x)$. Combining the two contributions
$\beta_1$ and $\beta_2$, we finally arrive at the compact expression
\begin{eqnarray}
&&\beta^\mu(x)
=
\Bigg[\,
\int\limits_{\Sigma}
d^3\vec\u\;
\big[\beta\cdot\opt_\u\big]
\nonumber\\
&&\quad
+\frac{1}{2}\sum_{\lambda,a}
\int\frac{d^3\k}{(2\pi)^3 2E_\k}
\int\limits_{\Sigma}
d^3\vec\u\, d^3\vec\v\;
\big[a_{-\k \lambda a}\cdot\opt_\u\big]
\big[a_{+\k \lambda a}\cdot\opt_\v\big]
\Bigg]{\cal A}^\mu(x)\; .
\label{eq:dA}
\end{eqnarray}

We can now use eqs.~(\ref{eq:green-NLO}) and (\ref{eq:dA})  to
obtain a compact expression for NLO corrections to ${\cal O}$ as 
\begin{eqnarray}
&&
{\cal O}_{_{\rm NLO}}(x,y)=
\Bigg[\,
\int\limits_{\Sigma}
d^3\vec\u
\big[\beta\cdot\opt_\u\big]
\nonumber\\
&&
\qquad+\frac{1}{2}\sum_{\lambda,a}
\int\frac{d^3\k}{(2\pi)^3 2E_\k}
\int\limits_{\Sigma}
d^3\vec\u\,d^3\vec\v\;
\big[a_{-\k \lambda a}\cdot\opt_\u\big]
\big[a_{+\k \lambda a}\cdot\opt_\v\big]
\Bigg]{\cal O}_{_{\rm LO}}[{\cal A}]
\nonumber\\
&&\qquad+\Delta{\cal O}_{_{\rm NLO}}(x,y)\; ,
\label{eq:O-NLO1}
\end{eqnarray}
where we recall that ${\cal O}_{_{\rm LO}}[{\cal A}]$ is the same
observable {\sl at leading order}, considered as a functional of the
value of the gauge fields on the initial surface $\Sigma$. The
corrective term $\Delta{\cal O}_{_{\rm NLO}}(x,y)$ is defined by 
\begin{eqnarray}
&&
\Delta{\cal O}_{_{\rm NLO}}(x,y)
\equiv
\frac{1}{2}
\sum_{\lambda,a}
\int\frac{d^3\k}{(2\pi)^3 2E_\k}\;
\int\limits_{\Sigma}
d^3\vec\u\, d^3\vec\v\nonumber\\
&&\qquad\times\left\{
\Big[
\big[a_{-\k \lambda a}\cdot\opt_\u\big]
{\cal A}^{i b}(x)
\Big]
\Big[
\big[a_{+\k \lambda a}\cdot\opt_\v\big]
{\cal A}^{j c}(y)
\Big]
\right.
\nonumber\\
&&
\qquad\quad
\left.
-
\Big[
\big[a_{+\k \lambda a}\cdot\opt_\u\big]
{\cal A}^{i b}(x)
\Big]
\Big[
\big[a_{-\k \lambda a}\cdot\opt_\v\big]
{\cal A}^{j c}(y)
\Big]
\right\}\; .
\label{eq:delta-O}
\end{eqnarray}
As we shall see later, this term $\Delta{\cal O}_{_{\rm NLO}}$ does not contain any large logarithm. 
Only the terms in the first and second lines of eq.~(\ref{eq:O-NLO1})
will be important for our later discussion of factorization.

\section{JIMWLK evolution for a single nucleus}
\label{sec:1nuc}
\begin{figure}[htbp]
\begin{center}
\resizebox*{!}{5cm}{\includegraphics{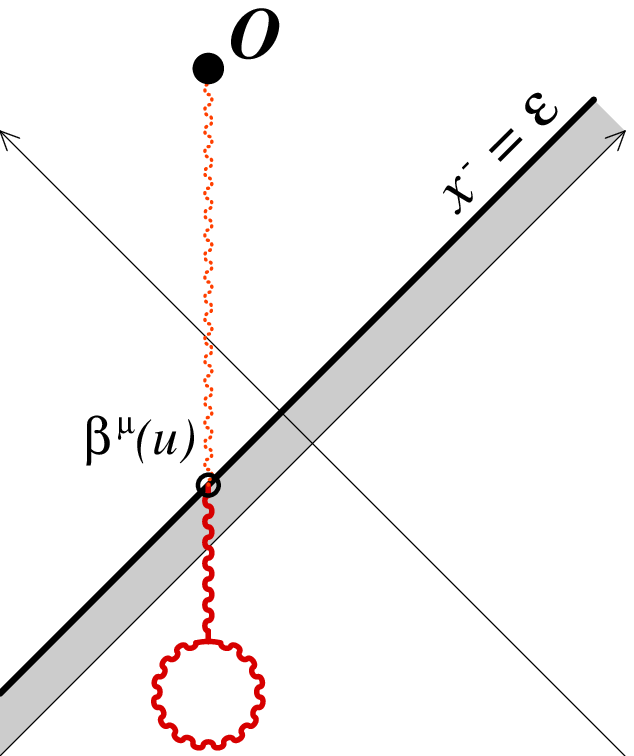}}
\hglue 10mm
\resizebox*{!}{5cm}{\includegraphics{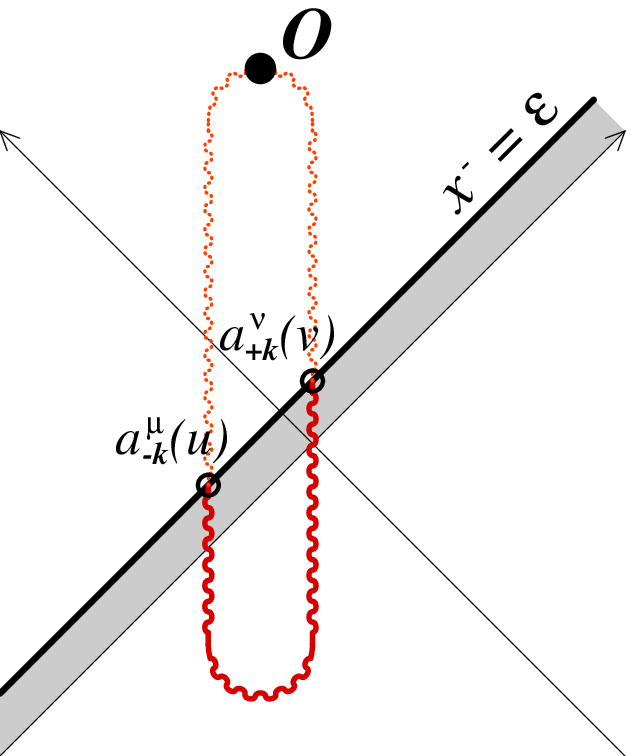}}
\end{center}
\caption{\label{fig:1nuc}NLO corrections in the single nucleus case,
  seen as an initial value problem on the surface $x^-=\epsilon$. The
  shaded area represents the domain where the nuclear color sources
  live ($0\le x^-\le\epsilon$). The field fluctuations represented in
  red continue to evolve in the region $x^->\epsilon$ until they hit
  the operator we want to evaluate. However, this evolution is
  entirely hidden in the dependence of the classical field upon its
  initial value at $x^-=\epsilon$, and we do not need to consider it
  explicitly.}
\end{figure}

Eq.~(\ref{eq:O-NLO1}) is central to our study of NLO corrections and
of factorization. In the rest of this section, we will show how this
formula is used to derive the JIMWLK evolution equation. In section
\ref{sec:2nuc}, we will show that it can be generalized to the
collision of two nuclei. A very convenient choice of initial surface $\Sigma$ in the derivation of the JIMWLK equation
 is the surface defined by
$x^-=\epsilon$. One should choose $\epsilon$ so that all the color
sources of the nucleus are located in the strip $0\le x^-\le\epsilon$.
An illustration of the objects involved in eq.~(\ref{eq:O-NLO1}) and
their localization in space-time is provided in figure \ref{fig:1nuc}.

\subsection{Gauge choice}
We need first to choose the gauge in which to perform this
calculation. Because the observable we wish to calculate and
everything else in eq.~(\ref{eq:O-NLO1}) is expressed in terms of
light cone gauge ($A^+=0$) quantities, we need to obtain $a_{\pm\k
  \lambda a}$ and $\beta$ in this gauge as well. However, as
previously mentioned, covariant current conservation is most easily
preserved in a gauge where the field fluctuations have no $-$
component. This is  because they do not induce a precession of the color
current $J^+$ while crossing the light cone. We are therefore going to
adopt the strategy advocated in
refs.~\cite{AyalaJMV1,AyalaJMV2,IancuLM1,FerreILM1}, that consists in
performing intermediate calculations in a gauge where $A^-=0$ and then gauge 
transforming the final result to $A^+=0$ gauge.

As discussed in detail in appendix \ref{app:green}, if one uses the LC
gauge and the surface $u^-=\epsilon$ as the initial surface, the
linear differential operator $a\cdot{\mathbbm T}_u$ that appears in
the identity (\ref{eq:aA}) should be defined as\footnote{We have
  omitted the color indices in this equation. $\Omega$ should be
  understood as a matrix in the $SU(N)$ group, and ${\cal A}$ as a
  column vector. $\Omega {\cal A}$ is therefore a column vector whose
  components are $(\Omega {\cal A})_c\equiv \Omega_{cb} {\cal A}_b$.}
\begin{eqnarray}
a\cdot{\mathbbm T}_u
&=&
\partial^-(\Omega(u)a^i(u))\frac{\delta}{\delta \big(\partial^-(\Omega(u){\cal A}^i(u))\big)}
+
\Omega(u)a^-(u)\frac{\delta}{\delta \big(\Omega(u){\cal A}^-(u)\big)}
\nonumber\\
&&
\qquad\qquad\qquad\qquad
+
\partial^\mu (\Omega(u)a_\mu(u))\frac{\delta}{\delta \big(\partial^\mu(\Omega(u){\cal A}_\mu(u))\big)}
\; ,
\label{eq:T-def}
\end{eqnarray}
where $\Omega$ is the adjoint color matrix\footnote{At first sight, $\Omega$ does not play any
role in the definition of ${\mathbbm T}_\u$ -- the necessity to
introduce this matrix $\Omega$ in the definition of ${\mathbbm T}_\u$
is also explained in the appendix \ref{app:green}.} that will be defined in
eq.~(\ref{eq:wilson}) . Note that this
operator in eq.~(\ref{eq:T-def}) contains a term for each of the field
components that must be specified on the initial surface to
know completely the field above this surface. This operator ${\mathbbm
  T}_u$ can therefore be interpreted as the generator of translations
of the initial condition for a classical solution of the Yang-Mills
equations. It is also important to note that the fluctuation field
$a^\mu(u)$ that multiplies this operator is evaluated just above the
initial surface (at $u^-= \epsilon$).  Therefore, because one does not
require its entire history beyond this surface, it can in general be
calculated analytically.

\subsection{Classical field}
\label{sec:classical}
Let us recall the structure of the classical background field itself.
As is well known, the field in the Lorenz gauge ($\partial_\mu
A^\mu=0$) has no $A^-$ component, and therefore fulfills the $A^-=0$
condition. Its explicit expression in terms of the color
source\footnote{The density $\rho$ of color sources is a gauge
  dependent quantity. When defined in the Lorenz gauge, we denote it
  with a tilde.} $\wt{\rho}$ in given by
\begin{equation}
\wt{\cal A}^+(x)=-\frac{1}{{\bs\partial}_\perp^2}\,\wt{\rho}(x^-,\x_\perp)\quad,\qquad
\wt{\cal A}^-=\wt{\cal A}^i=0\; .
\end{equation}
The gauge transformation that relates the classical background fields
in the $A^+=0$ gauge and the corresponding fields in Lorenz gauge
is\footnote{In this expression, $\Omega$ is a matrix in the {\sl
    group} $SU(N)$, while $\wt{\cal A}$ is a matrix in the adjoint
  representation of the {\sl algebra} $SU(N)$. The product
  $\Omega^\dagger \wt{\cal A}\Omega$ is a matrix in the $SU(N)$
  algebra.  Note that depending on the context we use the same symbol
  for an element ${\cal A}$ of the algebra (i.e. a matrix), and for
  the vector column made of its components ${\cal A}_c$ on the basis
  of the algebra. The relation between the two is of course ${\cal
    A}={\cal A}_c T^c$.}
\begin{equation}
{\cal A}^\mu=\Omega^\dagger \wt{\cal A}^\mu \Omega
+\frac{i}{g}\Omega^\dagger\partial^\mu \Omega\; ,
\label{eq:gt1}
\end{equation}
where the tilde denotes fields in the Lorenz gauge; those without a
tilde are in light cone gauge. Using the light cone gauge condition
$A^+=0$, we get
\begin{equation}
\partial^+\Omega=ig\wt{\cal A}^+\Omega\; ,
\label{eq:Omega0-1}
\end{equation}
which admits the Wilson line 
\begin{equation}
\Omega(x^-,\x_\perp)\equiv {\rm T}\,
\exp\Big[ig\int_{-\infty}^{x^-}dz^-\;\widetilde{\cal A}^+_a(z^-,\x_\perp)T^a\Big] 
\label{eq:wilson}
\end{equation}
as a solution. Note that because the color sources do not depend on $x^+$, $\wt{\cal
  A}^+$ and $\Omega$ depend only on $x^-$ and $\x_\perp$. The solution
of the classical equations of motion in light cone gauge is then
\begin{eqnarray}
&&
{\cal A}^+={\cal A}^-=0\; ,
\nonumber\\
&&
{\cal A}^i(x)=\frac{i}{g}
\Omega^\dagger(x^-,\x_\perp)\partial^i\Omega(x^-,\x_\perp)\; ,
\label{eq:class-LC}
\end{eqnarray}

We should comment here on the residual gauge freedom of the classical
solution. The most general solution of eq.~(\ref{eq:Omega0-1}) is
\begin{equation}
\Omega(x^-,\x_\perp)\Theta(x^+,\x_\perp)\; ,
\end{equation}
where $\Theta$ is an arbitrary $x^-$-independent gauge transformation.
With this more general choice, one obtains
\begin{eqnarray}
&&
{\cal A}^+=0\; ,
\nonumber\\
&&
{\cal A}^-=\frac{i}{g}\Theta^\dagger\partial^-\Theta\; ,
\nonumber\\
&&
{\cal A}^i=\Theta^\dagger\Big[\frac{i}{g}\Omega^\dagger\partial^i\Omega\Big]\Theta
+\frac{i}{g}\Theta^\dagger\partial^i\Theta\; .
\end{eqnarray}
The arbitrariness in the solution is because the condition $A^+=0$
does not fix completely the gauge and $x^-$-independent $\Theta$'s
span the residual gauge freedom.  Requiring that the classical gauge
field be of the form given in eq.~(\ref{eq:class-LC}) amounts to the
choice $\Theta\equiv {\bs 1}$. This choice is assumed in the rest of
this paper.

\subsection{Field fluctuations on the light cone}
To readers familiar with the structure of the JIMWLK Hamiltonian, the
structure of eq.~(\ref{eq:O-NLO1}) is already suggestive.  In the rest
of this section, we will show that the leading logarithmic
contributions in this formula -- terms that are linear in the
rapidity differences between the projectile and target relative to the
observed gluon -- can be absorbed into a redefinition of the
distribution of color sources of the nucleus. Our first task towards
this conclusion is to compute the value of the field fluctuations
$a_{\pm\k \lambda a}$ and $\beta$ just above the light cone on the
initial surface $u^-=\epsilon$.

Let us consider a small fluctuation $a^\mu$ on top of the classical
field ${\cal A}^\mu$. The relation between the two gauges must be
modified,
\begin{equation}
{\cal A}^\mu+a^\mu={\bar \Omega}^\dagger (\wt{\cal A}^\mu+\tilde{a}^\mu){\bar \Omega}
+\frac{i}{g}{\bar \Omega}^\dagger\partial^\mu {\bar \Omega}\; ,
\label{eq:gt2}
\end{equation}
with 
\begin{equation}
{\bar \Omega}\equiv(1+ig\omega)\Omega\; ,
\end{equation}
where $\omega$ has components of order unity. Using this ansatz in
eq.~(\ref{eq:gt2}), and keeping in mind that ${\cal A},\wt{\cal A}\sim
{\cal O}(g^{-1})$ while $a,\tilde{a}\sim {\cal O}(1)$, we obtain the relation 
\begin{equation}
a^\mu=\Omega^\dagger\Big(\tilde{a}^\mu-ig[\omega,\wt{\cal A}^\mu]-\partial^\mu\omega\Big)\Omega\; .
\label{eq:gt3}
\end{equation}
To determine $\omega$, as previously, apply the gauge condition
$a^+=0$. This gives  
\begin{equation}
\partial^+\omega+ig[\omega,\wt{\cal A}^+]=\tilde{a}^+\; ,
\label{eq:omega1-def}
\end{equation}
the solution of which can be written as
\begin{equation}
\omega(x)
=
\Omega(x^-,\x_\perp)f(x^+,\x_\perp)+
\int_{-\infty}^{x^-}dz^-
\;
\Omega(x^-,z^-;\x_\perp)\,
\tilde{a}^+(z^-,x^+,\x_\perp)\; .
\label{eq:omega1}
\end{equation}
In this equation $f$ is an arbitrary function that does not depend on
$x^-$, and $\Omega(x^-,z^-;\x_\perp)$ is an ``incomplete'' Wilson line
defined by
\begin{equation}
\Omega(x^-,z^-;\x_\perp)\equiv {\rm T}\,
\exp\Big\{ig\int_{z^-}^{x^-}dz^-\;\widetilde{\cal A}^+_a(z^-,\x_\perp)T^a\Big\}\; .
\label{eq:wilson1}
\end{equation}
The arbitrariness in the choice of the function $f_b$ again means that
there is a residual gauge freedom after we have imposed $a^+=0$. 

A crucial point in our derivation is how the residual gauge freedom is
fixed. We need small field fluctuations in order to represent the
propagators as in eqs.~(\ref{eq:G-+}) and (\ref{eq:G++}) as bi-linear
forms in these fluctuations. These equations are valid only if the
initial value of the fluctuations $a_{\pm\k \lambda a}$ are plane
waves with on-shell momenta; one can check easily that this is true
for the free propagators. Thus eq.~(\ref{eq:omega1}) must give plane
wave solutions for the field fluctuations in light cone gauge when
$x^-<0$.  This is simply achieved by taking plane waves for the
fluctuation $\tilde{a}^\mu$ in the original gauge and setting the
function $f$ to zero\footnote{We note that it is also possible to
  choose $\tilde{a}^\mu$'s that are not plane waves and a non-zero
  $f$ to achieve our requirement that $a^\mu$ be a plane wave.  This
  however makes the intermediate calculations more tedious.}.
Therefore, the requirement that eqs.~(\ref{eq:G-+}) and (\ref{eq:G++})
be valid leaves no residual gauge freedom.

We only need to know $\omega$ on our initial surface $\Sigma$ -- at
$x^-=\epsilon$. Because the components of $\Omega$ and of $\tilde{a}$
are all of order unity, it is legitimate to neglect the values of
$z^-$ that are between $0$ and $\epsilon$ in the integration in
eq.~(\ref{eq:omega1}). For $x^-=\epsilon$ and $z^-<0$, the incomplete
Wilson line is equal to the complete Wilson line (which has the lower
bound at $-\infty$). We therefore obtain
\begin{equation}
\omega(x^-=\epsilon)
=
\Omega(\x_\perp)
\int_{-\infty}^0 dz^-\;
\tilde{a}^{+}(z^-,x^+,\x_\perp)\; .
\label{eq:omega-LC}
\end{equation} 
Note also that when $\epsilon\le x^-$, the Wilson line becomes
independent of $x^-$ because all the color sources are in the strip
$0\le x^-\le \epsilon$. This explains why we only indicate $\x_\perp$
in its list of arguments.

Once $\omega$ has been determined, the $-$ and $i$ components of the
fluctuation in light cone gauge are determined from those in the
$A^-=0$ gauge to be
\begin{eqnarray}
a^-&=&\Omega^\dagger\Big(-\partial^-\omega\Big)\Omega\; ,
\nonumber\\
a^i&=&\Omega^\dagger\Big(\tilde{a}^i-\partial^i\omega\Big)\Omega\; .
\label{eq:rotated-field}
\end{eqnarray}
As we shall see shortly when we discuss the leading logarithmic
divergences, the only quantity we need is\footnote{Note that $(\Omega
  {\cal A} \Omega^\dagger)_c=\Omega_{cb} {\cal A}_b$, from the
  definition of the adjoint representation. With the notation where
  ${\cal A}$ is a column vector, this quantity would also be denoted
  by $(\Omega {\cal A})_c$.}
\begin{eqnarray}
\partial_\mu\big[\Omega a^\mu\Omega^\dagger\big]
&=&
\partial_\mu\big[\tilde{a}^\mu-\partial^\mu\omega-ig[\omega,\wt{\cal A}^\mu]\big]
\nonumber\\
&=&
-\partial^+ \partial^-\omega
-\partial^i\big[\tilde{a}^i-\partial^i\omega\big]\; ,
\label{eq:dmuamu1}
\end{eqnarray}
where we have used eq.~(\ref{eq:omega1-def}) and the fact that
$\tilde{a}^-=0$ in order to eliminate a few terms. Using the equation
for $\partial^+\omega$, as well as the fact that $\wt{\cal A}^+$ is
zero at $x^-=\epsilon$, we get
\begin{equation}
\partial_\mu\big[\Omega a^\mu\Omega^\dagger\big]
=
{\bs\partial}_\perp^2\omega
-\partial^-\tilde{a}^+
-\partial^i\tilde{a}^i\; .
\label{eq:dmuamu2}
\end{equation}

\def\EP{\bs\epsilon} Let us now consider specifically the fluctuations
$a_{\pm\k \lambda a}$. In the gauge $\tilde{a}^-=0$, their expression
below the light cone reads\footnote{Therefore, $\tilde{a}^{\mu
    b}_{\pm\k \lambda a}(x)=\tilde{\epsilon}^\mu_\lambda(\k)
  \delta^{ab} e^{\pm ik\cdot x}$.}
\begin{eqnarray}
\tilde{a}^\mu_{\pm\k \lambda a}(x)=\tilde{\epsilon}^\mu_\lambda(\k) T^a e^{\pm ik\cdot x}\; ,
\label{eq:planewave1}
\end{eqnarray}
with
\begin{eqnarray}
&&\tilde{\epsilon}^-_\lambda(\k)=0\; ,
\nonumber\\
&&
\sum_{\lambda=1,2}\tilde{\epsilon}^i_\lambda(\k)\tilde{\epsilon}^j_\lambda(\k)=-g^{ij}
\; ,
\nonumber\\
&&\tilde{\epsilon}^+_\lambda(\k)=\frac{\k_\perp\cdot\tilde{\EP}_{\lambda\perp}(\k)}{k^-}\; .
\end{eqnarray}
The formulas that govern the light cone crossing in this gauge have
been worked out in \cite{GelisM1}. Using these results, one finds the
following expressions for the fluctuation fields just above the
light cone:
\begin{eqnarray}
&&
\tilde{a}^{ib}_{\pm\k\lambda a}(x)=\Omega_{ba}(\x_\perp)\tilde{\epsilon}^i_\lambda(\k)
e^{\pm ik\cdot x}\;  ,
\nonumber\\
&&
\tilde{a}^{+b}_{\pm\k\lambda a}(x)=
\Big[\Omega_{ba}(\x_\perp)\tilde{\epsilon}^+_\lambda(\k)
\pm
\big(\partial^i\Omega_{ba}(\x_\perp)\big)\frac{1}{ik^-}\tilde{\epsilon}^i_\lambda(\k)\Big]e^{\pm ik\cdot x}\;  .
\end{eqnarray}
Note that for these field fluctuations, one has
\begin{equation}
\partial^-\tilde{a}^+_{\pm\k\lambda a}=\partial^i \tilde{a}^i_{\pm\k\lambda a}\; .
\end{equation}
Thus we have
\begin{eqnarray}
\partial_\mu\big[\Omega a^\mu_{\pm\k \lambda a}\Omega^\dagger\big]
&=&
\partial^i\big[\partial^i\omega -2 \tilde{a}^i_{\pm\k\lambda a}\big]\; .
\end{eqnarray}
Substituting eq.~(\ref{eq:planewave1}) in eq.~(\ref{eq:omega-LC}) gives the following expression for $\omega$
just above the light cone,
\begin{equation}
\omega_b = \mp 2i \Omega_{ba} \;\frac{k^j}{\k_\perp^2}\tilde{\epsilon}^j_\lambda(\k)\;e^{\pm i k\cdot x}\; .
\end{equation}
Therefore, 
\begin{equation}
\partial_\mu\big[\Omega a^\mu_{\pm\k \lambda a}\Omega^\dagger\big]_b
=-2\partial^i\Big[e^{\pm ik\cdot x}\,\alpha^{il b}_{\pm \k a}\,\epsilon^l_\lambda(\k)\Big]\; ,
\end{equation}
where we have introduced the shorthand notation 
\begin{eqnarray}
\epsilon_\lambda^l(\k)
&\equiv&
\Big(\delta^{lm}-2\frac{k^l k^m}{\k_\perp^2}\Big)\,\tilde{\epsilon}^m_{\lambda}(\k)\; ,
\nonumber\\
\alpha^{il b}_{\pm \k a}
&\equiv&
\Big(\delta^{il}-\frac{k^i k^l}{\k_\perp^2}\Big)\,\Omega_{ba}
\mp i \frac{k^l}{\k_\perp^2}\,\partial^i\Omega_{ba}\; .
\label{eq:alpha-def}
\end{eqnarray}

\subsection{Logarithmic divergences}
\label{sec:log}
Let us recall that our objective is to isolate the leading logarithmic
contributions to eq.~(\ref{eq:O-NLO1}). {}From the structure of this
equation, isolating these contributions requires that we examine
eq.~(\ref{eq:T-def}) term by term. As we shall see later, the
contribution in $\beta\cdot{\mathbbm T}$ (``virtual correction'') can
be derived from the term bilinear in ${\mathbbm T}$ (``real
correction''). Therefore, let us concentrate on the bilinear term for
now.

To determine the leading logarithmic contributions in the
real correction, we need to consider the integration over the on-shell
momentum $k^\mu$ as well.  It involves an integral
\begin{equation}
\int_0^{+\infty}
\frac{dk^+}{k^+}\; ,
\end{equation}
which potentially leads to logarithmic singularities both at $k^+\to
0$ and at $k^+\to +\infty$. Note that wherever $k^-$ appears in the
integrand, it should be replaced by the on-shell value
$k^-=\k_\perp^2/2k^+$. Inspecting the integrand of
eq.~(\ref{eq:O-NLO1}), one sees that the $k^+$ dependence contains
exponential factors
\begin{equation}
e^{i\frac{\k_\perp^2}{2k^+}(v^+-u^+)}\; .
\end{equation}
There is no factor depending on $v^--u^-$, because the points $u$ and
$v$ are both on the initial surface $\Sigma$, and thus have equal $-$
co-ordinates.  It is clear the integral converges at $k^+\to 0^+$
thanks to the oscillatory behavior of this exponential. On the other
hand, when $k^+\to +\infty$, the exponential goes to unity and one may
have a logarithmic singularity there. However, to truly have a
divergence, the other factors in the integrand should not have any
power of $1/k^+$.

Let us now examine these. The coefficients in the operator $a\cdot\opt_\u$ are the initial
values of $\Omega a^-,\partial^-(\Omega a^i)$ and $\partial_\mu
(\Omega a^\mu)$.  We need only to keep the coefficients that have no
power of $1/k^+$. One sees readily that this is not the case for $\Omega a^-$ or
$\partial^- (\Omega a^i)$~: these two quantities (compare
eq.~(\ref{eq:rotated-field}) to eqs.~(\ref{eq:dmuamu1}) 
and (\ref{eq:dmuamu2})) contain a factor $k^-\sim 1/k^+$. 

Thus,
as previously anticipated, the only divergence arises when one picks
up the term $\partial_\mu (\Omega a^\mu)$ both in $a\cdot\opt_\u$ and
$a\cdot\opt_\v$. 

In order to regularize the integral over $k^+$, we must introduce an
upper bound $\Lambda^+$. Physically, this cutoff is related to the
division of degrees of freedom one operates with in the CGC: the color
sources describe the fast partons and thus correspond to modes
$k^+>\Lambda^+$, while the fields represent the slow degrees of
freedom that have a longitudinal momentum $k^+<\Lambda^+$. Therefore,
when one performs a calculation in this effective description, the
longitudinal momentum of all the fields and field fluctuations should
not exceed $\Lambda^+$, in order not to overcount modes that are
already represented as part of the color sources $\rho$. The lower
scale in this logarithm is of the order of the longitudinal momentum
$p^+$ of the produced gluon. Therefore, the logarithm resulting from
the $k^+$ integration is a logarithm of $\Lambda^+/p^+$.

To pick up the logarithm, one should approximate the
exponential by unity. This implies that the coefficient of the
logarithm is independent of $u^+$ and $v^+$ or, in other words, it is
invariant under boosts in the $+z$ direction.  As we shall see, such
perturbations of $\partial_\mu (\Omega a^\mu)$ can be mapped to a
change in the color source $\tilde{\rho}$, and these logarithms can be
absorbed in a redefinition of the distribution $W[\tilde{\rho}]$.

\subsection{Real corrections}
Keeping only the term in $\partial_\mu (\Omega a^\mu)$ in
eq.~(\ref{eq:T-def}), and limiting ourselves to the divergent part of
the real correction for now, we see that we must evaluate the operator
\begin{eqnarray}
&&
\frac{1}{2\pi}\ln\left(\frac{\Lambda^+}{p^+}\right)
\int\frac{d^2\k_\perp}{(2\pi)^2}
\int d^2\u_\perp d^2\v_\perp
\nonumber\\
&&\qquad\times
\sum_a
\partial^i\Big(
\alpha^{ilb}_{-\k a}(\u_\perp)
e^{i\k_\perp\cdot\u_\perp}
\Big)
\partial^j\Big(
\alpha^{jlc}_{+\k a}(\v_\perp)
e^{-i\k_\perp\cdot\v_\perp}
\Big)
\nonumber\\
&&\qquad\times
\int du^+ dv^+
\frac{\delta^2}{\delta\partial_\mu \Omega(u)_{bd}
{\cal A}^\mu_d(u^+,\u_\perp)\;\;\delta\partial_\mu \Omega(v)_{ce}
{\cal A}^\mu_e(v^+,\v_\perp)}\; .
\label{eq:eta1}
\end{eqnarray}
Here, to avoid any confusion, we have written explicitly all the color
indices.  Note also that we have performed the sum over the two
polarization states of the field fluctuation in this
expression\footnote{A useful identity is\begin{equation*}
    \left(\delta^{il}-2\frac{k^i k^l}{\k_\perp^2}\right)
    \left(\delta^{lj}-2\frac{k^l k^j}{\k_\perp^2}\right)=\delta^{ij}\;
    .
\end{equation*}}.

The object on which this operator acts is the observable calculated at
leading order, considered as a functional of the initial value of the
fields ${\cal A}^i$ in light cone gauge. In this gauge, the
initial values of ${\cal A}^+$ and ${\cal A}^-$ are zero (provided the
residual gauge freedom is fixed as explained in section
\ref{sec:classical}). Moreover, from the set-up of the problem, it
turns out that these initial fields do not depend on $x^+$, 
\begin{equation}
{\cal A}^i(x^+,\x_\perp)={\cal A}^i(\x_\perp)\; ,
\end{equation}
and 
\begin{equation}
\partial_\mu \Omega(u)_{bd}
{\cal A}^\mu_d(u^+,\u_\perp)
=
-\partial^i\Omega(\u_\perp)_{bd}
{\cal A}^i_d(\u_\perp)\; .
\end{equation}
When we restrict ourselves to functionals that depend only on
$x^+$-independent initial fields, we can simply write\footnote{It is
  useful to recall that the dimension of a functional derivative
  operator is ${\rm Mass}^{-d({\cal A})-D}$ where $d({\cal A})$ is the
  mass dimension of the field with respect to which one is
  differentiating, and $D$ the mass dimension of the space in which
  this field lives. For instance
\begin{equation*}
  \frac{\delta}{\delta{\cal A}^i_b(u^+,\u_\perp)}\sim{\rm Mass}^{2}\quad,\qquad
  \frac{\delta}{\delta{\cal A}^i_b(\u_\perp)}\sim{\rm Mass}^{1}\; .
\end{equation*}}
\begin{equation}
\int du^+
\frac{\delta}{\delta\partial_\mu \Omega(u)_{bd}
{\cal A}^\mu_d(u^+,\u_\perp)}
=
-\frac{\delta}{\delta\partial^i\Omega(\u_\perp)_{bd}
{\cal A}^i_d(\u_\perp)}\; .
\label{eq:up-int}
\end{equation}

Our goal now is to relate the leading logarithmic contribution we have
identified to the JIMWLK evolution of the distribution of color
sources. As we have seen in the previous sections, the initial value
of the field in light cone gauge has a simple expression when
expressed in terms of the sources $\tilde{\rho}$ or fields $\wt{\cal
A}^+$ in Lorenz gauge. Therefore, we will try to make the connection
with the JIMWLK equation in this gauge. To do this, we must relate the
functional derivative $\delta/\delta\partial^i\Omega(\u_\perp)_{bd}
{\cal A}^i_d(\u_\perp)$ to the functional derivative
$\delta/\delta\wt{\cal A}^+$.  We begin by considering the light cone
gauge expression for the classical transverse gauge fields given by
eqs.~(\ref{eq:class-LC}) and (\ref{eq:wilson}). Rewriting ${\cal
A}^i(\x_\perp)$ more explicitly as
\begin{equation}
{\cal A}^i(x^-,\x_\perp)
=
-\int_{-\infty}^{x^-}dz^-\;
\Omega^\dagger(z^-,\x_\perp)
\Big(\partial^i\wt{\cal A}^+(z^-,\x_\perp)\Big)
\Omega(z^-,\x_\perp)\; ,
\end{equation}
one observes that a variation\footnote{It is natural that the size of
  the bin in which the field $\wt{\cal A}^+$ is changed plays a role
  here. Indeed, because $\wt{\cal A}^+$ is integrated over $x^-$ in
  the expression of ${\cal A}^i$, a change in a bin of zero width
  produces no change in ${\cal A}^i$. Note also that the factor $dx^-$
  in eq.~(\ref{eq:dA+-dAi}) is necessary on dimensional grounds.}
$\delta\wt{\cal A}^+(\epsilon,\x_\perp)$ of the field in covariant
gauge in the last $x^-$ bin (of width $dx^-$) leads to a change
$\delta{\cal A}^i(\x_\perp)$ of the initial value of the gauge field
in light cone gauge, given by
\begin{equation}
  \delta{\cal A}^i(\x_\perp)
  =
  -\Omega^\dagger(\x_\perp)
  \Big(\partial^i\delta\wt{\cal A}^+(\epsilon,\x_\perp)dx^-\Big)
  \Omega(\x_\perp)
  \; .
\label{eq:dA+-dAi}
\end{equation}
{}{}From this formula, we get the variation of
$\partial^i\Omega(\u_\perp)_{bd} {\cal A}^i_d(\u_\perp)$,
\begin{equation}
  \delta\Big[\partial^i\Omega(\u_\perp)_{bd} {\cal A}^i_d(\u_\perp)\Big]
  =
  -{\bs\partial}_\perp^2\delta\wt{\cal A}^+(\epsilon,\x_\perp)dx^-\; .
\end{equation}
Inverting this relation, one obtains 
\begin{equation}
\frac{\delta}{\delta\partial^i\Omega(\u_\perp)_{bd} {\cal A}^i_d(\u_\perp)}
=
-\int d^2\x_\perp
\;G(\u_\perp-\x_\perp)
\;
\frac{\delta}{\delta\wt{\cal A}^+_b(\epsilon_{_Y},\x_\perp)}\; .
\label{eq:dd}
\end{equation}
Here $G(\u_\perp-\x_\perp)$ is a two-dimensional propagator whose
main properties are discussed in appendix \ref{app:2d-prop}.  

It is important to observe that the functional derivatives on the left
and right hand side of this equation do not have the same dimensions.
This is because they are defined with respect to fields that live in
spaces with different dimensions. On the left hand side, the initial
transverse field in light cone gauge does not depend on $x^-$ as soon
as we are outside the nucleus and is therefore a function of
$\u_\perp$ only. On the right hand side, the field $\wt{\cal A}^+$
depends crucially on $x^-$. The $\epsilon_{_Y}$ argument in the right
hand side of eq.~(\ref{eq:dd}) is not integrated over, and should be
chosen as the value of $x^-$ where the last layer of quantum evolution
has produced its partons. This is the same as the location $\epsilon$
of the surface $\Sigma$ used for the initial conditions, but the
subscript $Y$ indicates that it may shift as the rapidity $Y$
increases.

We can now rewrite the operator in eq.~(\ref{eq:eta1}) as follows
\begin{eqnarray}
\frac{1}{2}
\ln\left(\frac{\Lambda^+}{p^+}\right)
\int d^2\x_\perp d^2\y_\perp\;
\eta^{bc}(\x_\perp,\y_\perp)
\frac{\delta^2}{\delta\wt{\cal A}^+_b(\epsilon_{_Y},\x_\perp)\delta\wt{\cal A}^+_c(\epsilon_{_Y},\y_\perp)}
\; ,
\label{eq:eta2}
\end{eqnarray}
where we have defined\footnote{We performed along the way an
integration by parts and used the identity in
eq.~(\ref{eq:deriv-2Dgreen}).}
\begin{eqnarray}
&&
\eta^{bc}(\x_\perp,\y_\perp)
\equiv
\frac{1}{4\pi^3}
\int\frac{d^2\k_\perp}{(2\pi)^2}\int d^2\u_\perp d^2\v_\perp
\sum_a
\alpha^{ilb}_{-\k a}(\u_\perp)\alpha^{jlc}_{+\k a}(\v_\perp)
\nonumber\\
&&\qquad\qquad\qquad\quad
\times e^{i\k_\perp\cdot(\u_\perp-\v_\perp)}\;
\frac{\u_\perp^i-\x_\perp^i}{(\u_\perp-\x_\perp)^2}\,
\frac{\v_\perp^j-\y_\perp^j}{(\v_\perp-\y_\perp)^2}\; .
\label{eq:eta3}
\end{eqnarray}
{}From eq.~(\ref{eq:alpha-def}), $\alpha^{ilb}_{\pm\k a}$ can
naturally be broken in two terms. If we keep only the first term in
each of the $\alpha$'s in eq.~(\ref{eq:eta3}), we obtain correspondingly 
\begin{eqnarray}
\eta^{bc}_{(1)}(\x_\perp,\y_\perp)
&=&
-\frac{1}{8\pi^4}
\int d^2\u_\perp d^2\v_\perp
\;
\frac{(\x_\perp^i-\u_\perp^i)}{(\x_\perp-\u_\perp)^2}
\;
\frac{(\y_\perp^j-\v_\perp^j)}{(\y_\perp-\v_\perp)^2}
\nonumber\\
&&\qquad\qquad\times
\Delta^{ij}(\u_\perp-\v_\perp)\,\Big[\Omega(u)\Omega^\dagger(v)-1\Big]_{bc}
\; .
\end{eqnarray}
Here the function $\Delta^{ij}$ is defined in eq.~(\ref{eq:dijG1}) of
appendix \ref{app:2d-prop}.  When we keep the first term in the first
$\alpha$ and the second term in the second $\alpha$ (or vice versa),
we get zero because the two terms in $\alpha$ are mutually orthogonal.
If we keep the second term in each of the $\alpha$'s, we obtain
\begin{eqnarray}
\eta^{bc}_{(2)}(\x_\perp,\y_\perp)
&=&
\frac{1}{\pi}
\int \frac{d^2\u_\perp}{(2\pi)^2}
\;
\frac{(\x_\perp^i-\u_\perp^i)(\y_\perp^i-\u_\perp^i)}{(\x_\perp-\u_\perp)^2(\y_\perp-\u_\perp)^2}\;
\nonumber\\
&&
\qquad\qquad
\times
\Big[
 \Omega(x)\Omega^\dagger(y)
-\Omega(x)\Omega^\dagger(u)
-\Omega(u)\Omega^\dagger(y)
+1\Big]_{bc}
\nonumber\\
&&
+\frac{1}{8\pi^4}\int d^2\u_\perp d^2\v_\perp\;
\frac{(\x_\perp^i-\u_\perp^i)}{(\x_\perp-\u_\perp)^2}
\;
\frac{(\y_\perp^j-\v_\perp^j)}{(\y_\perp-\v_\perp)^2}\;
\nonumber\\
&&\qquad\qquad\times
\Delta^{ij}(\u_\perp-\v_\perp)\;
\Big[\Omega(u)\Omega^\dagger(v)-1\Big]_{bc}\, .
\end{eqnarray}

When we add the two contributions, the terms involving
$\Delta^{ij}$ cancel, and we are finally left with
\begin{eqnarray}
\eta^{bc}(\x_\perp,\y_\perp)
&=&
\frac{1}{\pi}
\int \frac{d^2\u_\perp}{(2\pi)^2}
\;
\frac{(\x_\perp^i-\u_\perp^i)(\y_\perp^i-\u_\perp^i)}{(\x_\perp-\u_\perp)^2(\y_\perp-\u_\perp)^2}\;
\nonumber\\
&&
\quad
\times
\Big[
\Omega(x)\Omega^\dagger(y)
-\Omega(x)\Omega^\dagger(u)
-\Omega(u)\Omega^\dagger(y)
+1\Big]_{bc} .
\label{eq:eta-f}
\end{eqnarray}
This function is precisely the function $\eta^{bc}(\x_\perp,\y_\perp)$ that
appears in the JIMWLK equation~\cite{IancuLM1,FerreILM1}.

At this point, a word must be said of the term 
$\Delta{\cal O}_{_{\rm NLO}}$ in eq.~(\ref{eq:O-NLO1}). It is given by the difference of two terms that can be obtained from 
each other by exchanging $a_{+\k\lambda a}$ and $a_{-\k\lambda a}$.
Going back to the calculation of $\eta^{bc}(\x_\perp,\y_\perp)$, 
it is easy to check that for the calculation of the leading log term
these two terms give the same result and cancel. Physically this is due 
charge conjugation symmetry -- because the classical field is real we obtain 
the same result by exchanging the negative and positive energy 
asymptotic solutions for the quantum fluctuation, 
$\Delta{\cal O}_{_{\rm NLO}}$ is the difference between these
two and thus cancels out.

\subsection{Virtual corrections}
In the previous subsection, we focused on the real contribution to
eq.~(\ref{eq:O-NLO1}).  We now turn our attention to the term in
$\beta\cdot\opt_\u$ in eq.~(\ref{eq:O-NLO1}). Recall that
$\beta^\mu$ is the one-loop correction to the classical field in the
LC gauge and is evaluated in eq.~(\ref{eq:O-NLO1}) at $u^-=\epsilon$,
just above the region occupied by the nuclear sources. Mimicking the
evaluation of the real contribution, we can write
directly\footnote{One can confirm that $(\Omega)_{bd}\beta^-_d$ and
$\partial^-(\Omega)_{bd}\beta^i_d$ are zero and therefore cannot
appear in the operator $[\beta\cdot\opt_\u]$.}
\begin{eqnarray}
\label{eq:virtual-med}
&&
\int\limits_{u^-=\epsilon}du^+d^2\u_\perp\;
\big[\beta\cdot\opt_\u\big]
=
\nonumber\\
&&=
\int d^2\x_\perp\underbrace{\int d^2\u_\perp\;
G(\x_\perp-\u_\perp)\;
\partial_\mu^u\Big(\Omega(u)_{bd}\beta_d^\mu(u)\Big)}\;
\frac{\delta}{\delta\wt{\cal A}_b^+(\epsilon_{_Y},\x_\perp)}\; .
\\
&&\hskip 34mm
\ln\Big(\frac{\Lambda^+}{p^+}\Big)\,\nu^b(\x_\perp)
\nonumber
\end{eqnarray}
We anticipate that a large logarithm in the $k^+$ integral will show
up in this quantity, and we have defined $\nu^b(\x_\perp)$ as its
coefficient.  Note that in this definition of the function
$\nu^b(\x_\perp)$, the value\footnote{The value of $u^+$ is irrelevant
because the 1-point function $\beta^\mu(u)$ propagating over an
$x^+$-independent background field (and with a vanishing initial
condition in the past) is independent of $u^+$.} of $u^-$ is
$u^-=\epsilon$.

We begin with the Green's formula for the 1-point
function $\beta^\mu(u)$, where the initial surface is taken at $v^-=0$
(instead of $v^-=\epsilon$),
\begin{eqnarray}
\beta^\mu(u)
&=&
\int\limits_{v^->0} d^4v\; D_{0,_R}^{\mu\nu}(u,v)\;
\Big[
\frac{\partial^2 U({\cal A})}{\partial{\cal A}^\nu(v)\partial {\cal A}^\rho(v)}
\;\beta^\rho(v)
\nonumber\\
&&\qquad\qquad
+
\frac{1}{2}
\frac{{\partial^3 U({\cal A})}}
{\partial {\cal A}^\nu(v)\partial {\cal A}^\rho(v) \partial {\cal A}^\sigma(v)}\,
{\cal G}^{\rho\sigma}_{++}(v,v)\Big]
\; .
\end{eqnarray}
By this choice of the initial surface, we do not have a boundary term,
because $\beta^\mu$ is zero at $u^-\le 0$. The propagator ${\cal
  G}^{\rho\sigma}_{++}(v,v)$ can be expressed in terms of the field
fluctuations $a_{\pm \k\lambda a}$ by using eq.~(\ref{eq:G++}).
Consider now the Green's formula for the fluctuation $a_{+ \k\lambda
  a}$ we introduced in eq.~(\ref{eq:tmp4-green}), but written this
time for an initial surface at $u^-=0$,
\begin{equation}
a_{+ \k\lambda a}^\mu(x)
=
\int\limits_{y^->0}d^4y\;
D_{0,_R}^{\mu\nu}(x,y)\;
\frac{\partial^2 U({\cal A})}{\partial{\cal A}^\nu(y)\partial {\cal A}^\rho(y)}
\;a_{+ \k\lambda a}^\mu(y)
+{\cal B}^\mu_0[a_{+ \k\lambda a}]\; .
\end{equation}
In this formula, both the fluctuation $a_{+ \k\lambda a}$ and the
derivative of the gauge potential depend on the background classical
field in LC gauge. Let us apply to this equation the
operator\footnote{This operator is similar to the operator
$a\cdot\opt$ previously defined, but it performs the
replacement of fields inside the region of the sources, instead of
just on the surface of this region.} $\big[a_{-\k\lambda a}\cdot{\cal
T}\big]$ that substitutes one power of the background field by a power
of $a_{-\k\lambda a}$. By defining
\begin{eqnarray}
\xi^\mu(u)\equiv
\frac{1}{2}\sum_{\lambda,a}
\int\frac{d^3\k}{(2\pi)^32E_\k}\int\limits_{v^->0}d^4v\;
\big[a_{-\k\lambda a}\cdot{\cal T}_\v\big]\;
a_{+ \k\lambda a}^\mu(u)
\; ,
\end{eqnarray}
we obtain for this object the Green's formula
\begin{eqnarray}
\xi^\mu(u)
&=&
\int\limits_{v^->0} d^4v\; D_{0,_R}^{\mu\nu}(u,v)\;
\Big[
\frac{\partial^2 U({\cal A})}{\partial{\cal A}^\nu(v)\partial {\cal A}^\rho(v)}
\;\xi^\rho(v)
\nonumber\\
&&\qquad\qquad
+
\frac{1}{2}
\frac{{\partial^3 U({\cal A})}}
{\partial {\cal A}^\nu(v)\partial {\cal A}^\rho(v) \partial {\cal A}^\sigma(v)}\,
{\cal G}^{\rho\sigma}_{++}(v,v)\Big]
\; ,
\end{eqnarray}
where we used eq.~(\ref{eq:G++}) for the propagator that appears
in the source term. We see that $\xi^\mu$ and $\beta^\mu$ are 
identical. Therefore, we have proved that
\begin{eqnarray}
\beta^\mu_d(u)\equiv
\frac{1}{2}\sum_{\lambda,a}
\int\frac{d^3\k}{(2\pi)^32E_\k}\int\limits_{v^->0}d^4v\;
\big[a_{-\k\lambda a}\cdot{\cal T}_\v\big]\;
a_{+ \k\lambda a}^{\mu d}(u)
\; .
\end{eqnarray}

\begin{figure}[htbp]
\begin{center}
\resizebox*{!}{5cm}{\includegraphics{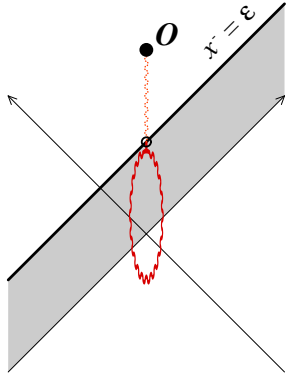}}
\end{center}
\caption{\label{fig:tadpole-LLog}Leading logarithmic contribution of
the tadpole diagram.}
\end{figure}

Inserting this expression into the definition of $\nu^b(\x_\perp)$, we
obtain
\begin{eqnarray}
&&\ln\Big(\frac{\Lambda^+}{p^+}\Big)\,\nu^b(\x_\perp)
=
\frac{1}{2}\sum_{\lambda,a}
\int\frac{d^3\k}{(2\pi)^32E_\k}\int\limits_{v^->0}d^4v\;
\big[a_{-\k\lambda a}\cdot{\cal T}_\v\big]
\nonumber\\
&&\quad
\times\int d^2\u_\perp\;
G(\x_\perp-\u_\perp)\;
\partial_\mu^u\Big(\Omega(u)_{bd}a_{+ \k\lambda a}^{\mu d}(u)\Big)\; .
\label{eq:nu-med}
\end{eqnarray}
To obtain a divergence at large $k^+$, we need to tame the
oscillations in this variable which exist because we have now
$u^-=\epsilon$ while $v^-$ can be anywhere in the range
$[0,\epsilon]$. These oscillations are damped only if $v^-$ is in the
immediate vicinity of $u^-=\epsilon$. As a corollary, note that the
left diagram in figure \ref{fig:1nuc} is therefore a bit misleading
because the tadpole contribution depicted vanishes when the upper
vertex of the tadpole is below the light cone.  In fact, to have a
leading logarithmic contribution, this vertex of the tadpole must be
very close to the surface $u^-=\epsilon$, as illustrated in figure
\ref{fig:tadpole-LLog}.

For sufficiently small $dx^-$, we can use
\begin{equation}
\lim_{dx^-\to 0}\int\limits_{\epsilon-dx^-}^\epsilon dv^-\;
\big[a_{-\k\lambda a}\cdot{\cal T}_\v\big]
=
a_{-\k\lambda a}\cdot\opt_\v\; ,
\end{equation}
namely, we recover the operator that substitutes the background field
by the fluctuation in the last layer at $v^-=\epsilon$. Again, using
the eqs.~(\ref{eq:up-int}) and (\ref{eq:dd}) from the previous
subsection, we obtain the operator
\begin{eqnarray}
&&
\int\limits_{v^-=\epsilon}dv^+d^2\v_\perp\;
\big[a_{-\k\lambda a}\cdot\opt_\v\big]
\empile{=}\over{\rm LLog}
\int d^2\y_\perp
\int d^2\v_\perp\;
G(\y_\perp-\v_\perp)\;\nonumber \\
&&\qquad\qquad\qquad\qquad
\times\; \partial_\nu^v\Big(\Omega(v)_{ce}a_{-\k\lambda a}^{\nu e}(v)\Big)\;
\frac{\delta}{\delta\wt{\cal A}_c^+(\epsilon_{_Y},\y_\perp)}\; .
\end{eqnarray}
When inserted in eq.~(\ref{eq:nu-med}), this gives
\begin{eqnarray}
&&\ln\Big(\frac{\Lambda^+}{p^+}\Big)\,\nu^b(\x_\perp)
=
\frac{1}{2}\int d^2\y_\perp\;
\underline{\sum_{\lambda,a}
\int\frac{d^3\k}{(2\pi)^32E_\k}}
\nonumber\\
&&\qquad\qquad\times
\underline{\int d^2\v_\perp\;
G(\y_\perp-\v_\perp)\;
\partial_\nu^v\Big(\Omega(v)_{ce}a_{-\k\lambda a}^{\nu e}(v)\Big)}
\nonumber\\
&&\qquad\qquad\times
\frac{\delta}{\delta\wt{\cal A}_c^+(\epsilon_{_Y},\y_\perp)}\;
\underline{
\int d^2\u_\perp\;
G(\x_\perp-\u_\perp)\;
\partial_\mu^u\Big(\Omega(u)_{bd}a_{+ \k\lambda a}^{\mu d}(u)\Big)}\; .
\nonumber\\
&&
\end{eqnarray}
Note that the product of the underlined terms, by themselves are just 
\begin{equation}
\ln\Big(\frac{\Lambda^+}{p^+}\Big)\,
\eta^{bc}(\x_\perp,\y_\perp)\; .
\end{equation}
The final step in our derivation is to note that when $\wt{\cal A}$
shares a color index with $\Omega$, we have the
identity~\cite{IancuLM1,FerreILM1,Muell4}
\begin{equation}
\frac{\delta}{\delta\wt{\cal A}_c^+(\epsilon_{_Y},\y_\perp)}\;
\partial_\nu^v\Big(\Omega(v)_{ce}a_{-\k\lambda a}^{\nu e}(v)\Big)
=0\; ,
\end{equation}
because of the antisymmetry of the adjoint generators of
$SU(N)$. We can therefore move the operator ${\delta}/{\delta\wt{\cal
A}_c^+(\epsilon_{_Y},\y_\perp)}$ immediately after the measure
$d^2\y_\perp$ to obtain 
\begin{equation}
\nu^b(\x_\perp)=
\frac{1}{2}
\int d^2\y_\perp\;
\frac{\delta}{\delta\wt{\cal A}_c^+(\epsilon_{_Y},\y_\perp)}\;
\eta^{bc}(\x_\perp,\y_\perp)\; ,
\label{eq:nu-final}
\end{equation}
which is identical to the relation between $\eta^{bc}$ and $\nu^b$ in
the JIMWLK equation.

\subsection{JIMWLK equation}
We shall now combine the real and virtual corrections to write
the JIMWLK equation. Using the real correction in eq.~(\ref{eq:eta2})
and the virtual one given by eqs.~(\ref{eq:virtual-med}) 
and (\ref{eq:nu-final}) we can write
the total NLO correction, eq.~(\ref{eq:O-NLO1}), in the form
\begin{equation}
{\cal O}_{_{\rm NLO}}
\empile{=}\over{\rm LLog}
\ln\left(\frac{\Lambda^+}{p^+}\right)\;{\cal H}
\,{\cal O}_{_{\rm LO}}[\wt{\cal A}^+]
\label{eq:O-NLO3}
\end{equation}
where we have introduced the JIMWLK Hamiltonian,
\begin{equation}
{\cal H}
\equiv \frac{1}{2}\int d^2\x_\perp d^2\y_\perp
\frac{\delta}{\delta\wt{\cal A}_c^+(\epsilon_{_Y},\y_\perp)}\;
\eta^{bc}(\x_\perp,\y_\perp)\;
\frac{\delta}{\delta\wt{\cal A}_b^+(\epsilon_{_Y},\x_\perp)}\; .
\label{eq:H-JIMWLK}
\end{equation}
Although the coupling does not appear explicitly in the Hamiltonian,
it is of order $\as$ because of the presence of two functional
derivatives with respect to classical fields that are of order
$g^{-1}$.

We noted that the observable ${\cal O}$ at leading order can be expressed as 
a functional of the classical gauge field $\wt{\cal A}^+$ in covariant
gauge. The average of this observable over all the
configurations of the field $\wt{\cal A}^+$, up to NLO, can be expressed as 
\begin{equation}
\left<{\cal O}_{_{\rm LO}}+{\cal O}_{_{\rm NLO}}\right>
\equiv
\int \big[D\wt{\cal A}^+\big]\,W\big[\wt{\cal A}^+\big]\,
\Big[{\cal O}_{_{\rm LO}}+{\cal O}_{_{\rm NLO}}\Big]\; .
\end{equation}
At the leading logarithmic level, this can be rewritten as
\begin{equation}
\left<{\cal O}_{_{\rm LO}}+{\cal O}_{_{\rm NLO}}\right>
\empile{=}\over{\rm LLog}
\int \big[D\wt{\cal A}^+\big]\,
\Big\{\Big[1+\Delta Y\,{\cal H}\Big]W\big[\wt{\cal A}^+\big]\Big\}\,
{\cal O}_{_{\rm LO}}[\wt{\cal A}^+]\; ,
\label{eq:O-NLO5}
\end{equation}
where we denote $\Delta Y\equiv \ln(\Lambda^+/p^+)$. Note that $\Delta
Y$ is also the rapidity interval between the slowest incoming sources
(that have $k^+\sim\Lambda^+$) and the measured gluon. To obtain this
equation, one uses the Hermiticity of ${\cal H}$ with respect to the
functional integration over $\wt{\cal A}^+$.  In writing this
equation, we have absorbed all the leading logarithms of $k^+$ into a
redefinition of the distribution $W\big[\wt{\cal A}^+\big]$,
\begin{equation}
W\big[\wt{\cal A}^+\big]
\quad
\to
\quad
\Big[1+\Delta Y\,{\cal H}\Big]W\big[\wt{\cal A}^+\big]\; .
\end{equation}
This suggests that the distribution $W\big[\wt{\cal A}^+\big]$ should
depend on the scale $\Lambda^+$ that separates the modes described as
static sources from the modes described as dynamical fields in the CGC
description. Of course, this is not surprising in an effective theory
based on such a separation of the degrees of freedom.  For this
reason, it should be denoted as $W_{_{\Lambda^+}}[\wt{\cal A}^+]$.
Therefore eq.~(\ref{eq:O-NLO5}) can be written as
\begin{equation}
\left<{\cal O}_{_{\rm LO}}+{\cal O}_{_{\rm NLO}}\right>
\empile{=}\over{\rm LLog}
\int \big[D\wt{\cal A}^+\big]\,
\Big\{\Big[1+\ln\left(\frac{\Lambda^+}{p^+}\right){\cal H}\Big]
\,W_{_{\Lambda^+}}\big[\wt{\cal A}^+\big]\Big\}\;
{\cal O}_{_{\rm LO}}[\wt{\cal A}^+]\; .
\label{eq:O-NLO6}
\end{equation}
Because $\Lambda^+$ is a an unphysical separation scale, the
expectation value of observables should not depend on this parameter.
Differentiating the previous equation with respect to $\Lambda^+$ and
requiring that the r.h.s be zero, we get\footnote{To avoid confusion,
  recall that ${\cal H}$, and hence $\partial W/\partial{\Lambda^+}$,
  are of order $\as$. Therefore, for consistency, one should not
  keep the term proportional to ${\cal H}(\partial
  W/\partial{\Lambda^+})$ because it is of order $\as^2$ and
  therefore beyond the accuracy of the present calculation.}
\begin{equation}
\frac{\partial }{\partial \ln(\Lambda^+)}W_{_{\Lambda^+}}[\wt{\cal A}^+]
=
-{\cal H}\,W_{_{\Lambda^+}}[\wt{\cal A}^+]\; .
\end{equation}
Equivalently, if $Y\equiv\ln(P^+/\Lambda^+)$ denotes the rapidity
separation between the fragmentation region of the nucleus (located at
$k^+\sim P^+$) and the rapidity down to which partons are described as
static color sources, we have
\begin{equation}
\frac{\partial }{\partial Y}W_{_Y}[\wt{\cal A}^+]
=
{\cal H}\;W_{_Y}[\wt{\cal A}^+]\; ,
\label{eq:JIMWLK}
\end{equation}
which is the JIMWLK equation that drives the $Y$
dependence of the distribution $W_{_Y}[\wt{\cal A}^+]$. 

The above considerations also indicate that the distribution
$W_{_Y}[\wt{\cal A}^+]$ must be evolved to a scale $\Lambda^+$
comparable to the typical longitudinal momentum in the observable to
avoid large residual logs contributing to the latter. Therefore, at
leading logarithmic accuracy, the expectation value of the observable
is given by
\begin{equation}
\left<{\cal O}\right>_{\rm LLog}=
\int\big[D\wt{\cal A}^+\big]\,W_{_Y}\big[\wt{\cal A}^+\big]\;
{\cal O}_{_{\rm LO}}[\wt{\cal A}^+]\; ,
\label{eq:O-NLO7}
\end{equation}
with $Y=\ln(P^+/p^+)$ the rapidity separation between the beam and the
observable and $W_{_Y}\big[\wt{\cal A}^+\big]$ given by the solution
of eq.~(\ref{eq:JIMWLK}).

\subsection{All order resummation of leading logs}
Thus far, we only considered 1-loop corrections that generate one
power of the large logarithm of $P^+$. On this basis, we deduced an
evolution equation for $W[\wt{\cal A}^+]$ using renormalization group
arguments. However, the solution of the RG equation is equivalent to a
resummation of all $n$-loop diagrams that have $n$ powers of large
logarithms of $p^+$. We shall here analyze the structure of higher
loop contributions to confirm whether the all loop resummation
performed by the RG equation is justified.

We will not perform here a detailed analysis of these leading $n$-loop
graphs to show that we indeed recover the solution of
eq.~(\ref{eq:JIMWLK}). More modestly, we will work {\it a posteriori}
by examining the solution of the JIMWLK equation to see what the
$n$-loop graphs that it resums are. Before proceeding, it is useful to
recall a crucial property of the JIMWLK Hamiltonian defined in
eq.~(\ref{eq:H-JIMWLK}). The operator ${\cal H}$ contains derivatives
with respect to the field $\wt{\cal A}^+(\epsilon_{_Y},\x_\perp)$ and
its coefficients depend on all the fields $\wt{\cal
  A}^+(x^-,\x_\perp)$ for $0\le x^-\le \epsilon_{_Y}$. For this
reason, we will denote it ${\cal H}(Y)$, where the endpoint
$\epsilon_{_Y}$ at which the derivatives act is related to $Y$ by
$Y\sim\ln(\epsilon_{_Y})$.  It is important to note that in a product
${\cal H}(y_1){\cal H}(y_2)$, the derivatives in ${\cal H}(y_1)$ do
not act on the coefficients of ${\cal H}(y_2)$ if $y_1>y_2$.

The JIMWLK equation should now be written as
\begin{equation}
\frac{\partial }{\partial Y}W_{_Y}[\wt{\cal A}^+]
=
{\cal H}(Y)\,W_{_Y}[\wt{\cal A}^+]\; ,
\label{eq:JIMWLK1}
\end{equation}
and its solution reads
\begin{equation}
W_{_Y}[\wt{\cal A}^+]
={\cal U}(Y)\;
W_0[\wt{\cal A}^+]\; ,
\label{eq:JIMWLK-sol}
\end{equation}
with
\begin{equation}
{\cal U}(Y)\equiv T_{_Y}\left[\exp\,\int_0^Y dy\; {\cal H}(y)\right]\; .
\end{equation}
In this equation, $T_{_Y}$ denotes a ``rapidity ordering'' such that
products of ${\cal H}$'s in the Taylor expansion of the exponential
are ordered from left to right in order of decreasing $y$.
$W_0[\wt{\cal A}^+]$ is a non-perturbative initial condition.  ${\cal
  U}(Y)$ is the evolution operator for the Hamiltonian ${\cal H}(Y)$.
Inserting eq.~(\ref{eq:JIMWLK-sol}) into eq.~(\ref{eq:O-NLO7}), we get
\begin{equation}
\left<{\cal O}\right>_{\rm LLog}=
\int\big[D\wt{\cal A}^+\big]\,W_0\big[\wt{\cal A}^+\big]\,
{\cal U}^\dagger(Y)\;
{\cal O}_{_{\rm LO}}[\wt{\cal A}^+]\; .
\label{eq:O-NLO8}
\end{equation}
Because ${\cal H}(y)$ is Hermitian, the Hermitian conjugate of the
evolution operator ${\cal U}(Y)$ is the same operator with the 
rapidity ordering reversed~: 
\begin{equation}
{\cal U}^\dagger(Y)\equiv \overline{T}_{_Y}\left[\exp\,\int_0^Y dy\;
{\cal H}(y)\right]\; ,
\end{equation}
where $\overline{T}_{_Y}$ denotes the anti-rapidity ordering. The
expansion of ${\cal U}^\dagger$ to order one in ${\cal H}$ gives the
leading logarithmic one-loop contributions that we have evaluated
earlier in this section. (See eq.~(\ref{eq:O-NLO5}) for instance.) 

\begin{figure}[htbp]
\begin{center}
\resizebox*{!}{4cm}{\includegraphics{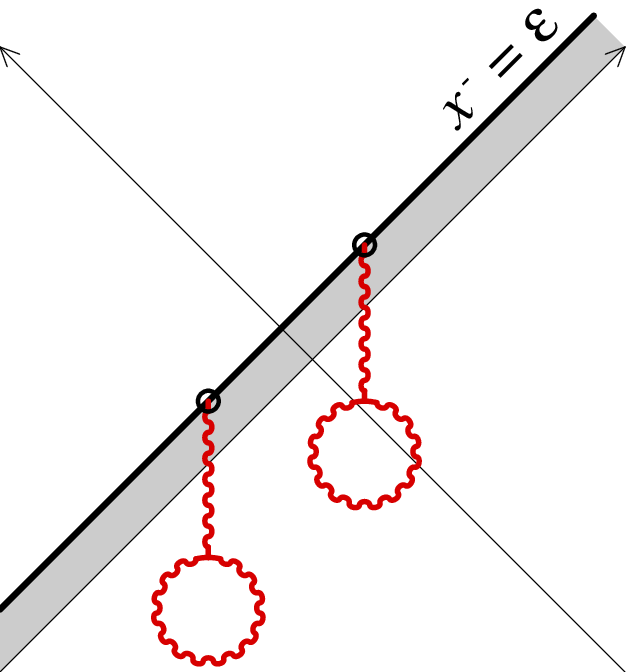}}
\hglue 5mm
\resizebox*{!}{4cm}{\includegraphics{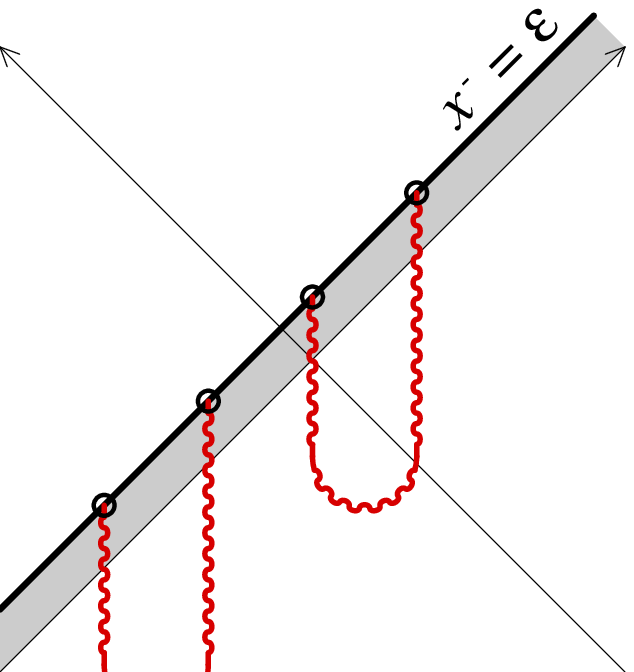}}
\hglue 5mm
\resizebox*{!}{4cm}{\includegraphics{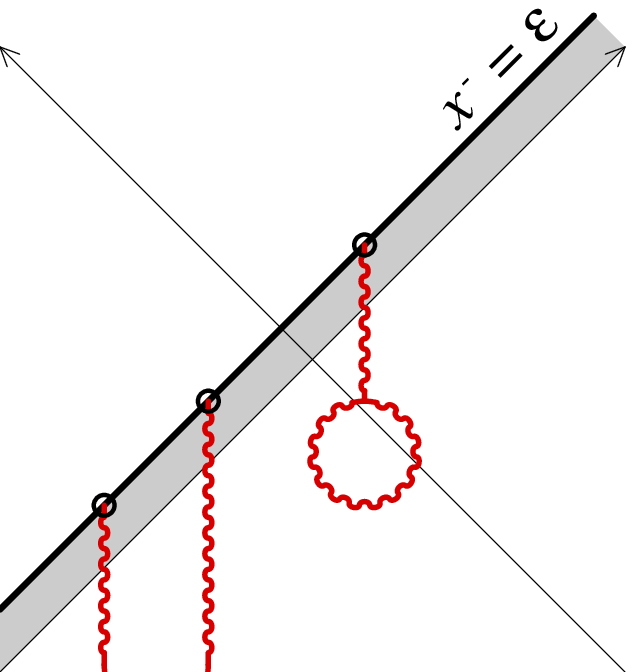}}
\end{center}
\caption{\label{fig:2loop-2}2-loop contributions made of products of
pieces already encountered at 1-loop. Although we do not make this
distinction in the figure, one of the factors is attached at a
slightly smaller value of $x^-$, because the two Hamiltonians in
eq.~(\ref{eq:tmp1}) are at different rapidities.}
\end{figure}

If one expands it to second order, we see
that the leading logarithmic contributions in the observable at two
loops should be given by
\begin{equation}
{\cal O}_{_{\rm NNLO}}\empile{=}\over{\rm LLog}
\int_0^Y dy_1 \int_0^{y_1} dy_2\;{\cal H}(y_2)\,{\cal H}(y_1)\;{\cal O}_{_{\rm LO}}[\wt{\cal A}^+]\; .
\label{eq:tmp1}
\end{equation}
Because $y_2<y_1$, the derivatives in ${\cal H}(y_2)$ can act on the
coefficients $\eta$ and $\nu$ of ${\cal H}(y_1)$. Let us first
consider the terms where this does not happen, namely where the
derivatives in ${\cal H}(y_2)$ act directly on ${\cal O}_{_{\rm
LO}}[\wt{\cal A}^+]$. These terms correspond to the graphs depicted in
figure \ref{fig:2loop-2}. If we look only at what happens below the
line $x^-=\epsilon$, these contributions are just disconnected
products of terms we had already at 1-loop.
The analysis we performed of the logarithmic contributions at one
loop extends trivially to these terms and it is easy to see that they
have two powers of the logarithms.

\begin{figure}[htbp]
\begin{center}
\resizebox*{!}{4cm}{\includegraphics{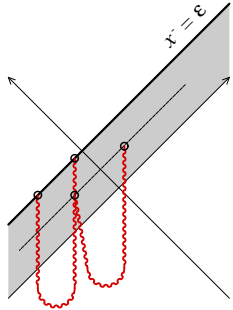}}
\end{center}
\caption{\label{fig:2loop-3}Example of term obtained when the
derivatives in ${\cal H}(y_2)$ can act on the coefficients of ${\cal
H}(y_1)$. Here, one of the derivatives in ${\cal H}(y_2)$ acts of the
function $\eta$ of ${\cal H}(y_1)$ and the second derivative in ${\cal
H}(y_2)$ acts on ${\cal O}_{_{\rm LO}}$.}
\end{figure}

In addition, eq.~(\ref{eq:tmp1}) also contains terms in which at least
one of the derivatives in ${\cal H}(y_2)$ acts on the coefficients of
${\cal H}(y_1)$. This corresponds to topologies of the type displayed
in figure \ref{fig:2loop-3}.
Such terms, that have a gluon vertex inside the region where the
sources live, have a large logarithm for the same reason that the
tadpole has a logarithm in the 1-loop terms. Thus one can see that it
is crucial to properly order the powers of the Hamiltonian
${\cal H}$ in rapidity not to lose these terms\footnote{For instance, if
the ordering of the two Hamiltonians in eq.~(\ref{eq:tmp1}) is
reversed, we get only the terms of figure \ref{fig:2loop-2}.}.

\begin{figure}[htbp]
\begin{center}
\resizebox*{!}{4cm}{\includegraphics{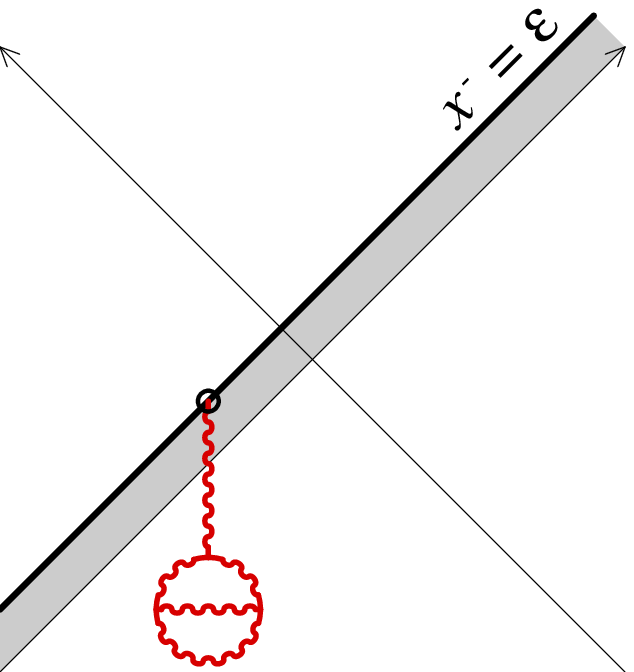}}
\hglue 5mm
\resizebox*{!}{4cm}{\includegraphics{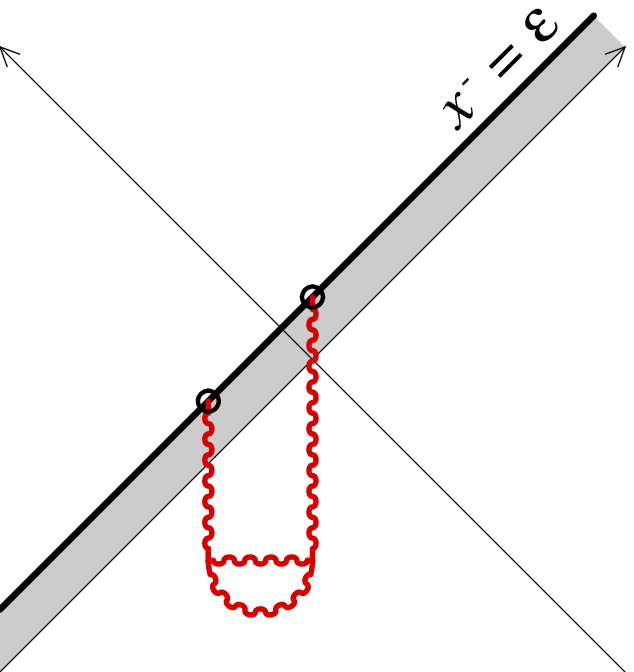}}
\end{center}
\caption{\label{fig:2loop-1}Some of the 2-loop corrections to the
observable ${\cal O}$ that do not appear at leading log.}
\end{figure}

Finally, there also exist at two loops some topologies that never
appear in eq.~(\ref{eq:tmp1}), such as those of figure
\ref{fig:2loop-1}.
The contributions in this figure are 1-loop corrections to the
coefficients of the operators $\opt_{\u,\v}$ in eq.~(\ref{eq:O-NLO1}).
In other words, these terms generate corrections of order $\as$
to the coefficients in the JIMWLK equation, and do not have double
logs of $\Lambda^+$. This explains why they are not generated by the
leading log formula in eq.~(\ref{eq:tmp1}).

\section{Nucleus-nucleus collisions}
\label{sec:2nuc}
In the previous section, we obtained an expression for resummed
leading logarithmic inclusive gluon observables in a single nucleus in
terms of the equivalent leading order observable. Along the way, we
presented a novel derivation of the JIMWLK evolution equation. In this
section, we will extend our analysis to the case of nuclear
collisions.  We will show that the leading logarithms of $k^\pm$ that
arise in the calculation of loop corrections to the single inclusive
gluon spectrum can be factored out in the distributions $W[\wt{\cal
  A}^+_1]$ and $W[\wt{\cal A}^-_2]$ that describe the two incoming
nuclei. This result will complete a proof of factorization of leading
logarithms of $1/x_{1,2}$ for inclusive observables in nucleus-nucleus
collisions.

\subsection{Inclusive observables at leading order}
As in the single nucleus case, our discussion is valid for an
inclusive multi-gluon operator ${\cal O}$ (corresponding to a moment
of the multiplicity or energy distribution produced in nucleus-nucleus
collisions) but for simplicity, we will focus on the first moment of
the multiplicity distribution -- the inclusive gluon spectrum.  As we discussed in \cite{GelisV2,GelisLV2}, 
the inclusive single particle spectrum in nucleus-nucleus collisions can be expressed as
\begin{eqnarray}
E_\p\frac{dN}{d^3\p}&=&\frac{1}{16\pi^3}
\lim_{x_0\to+\infty}\int d^3\x d^3\y
\;e^{ip\cdot(x-y)}
\;(\partial_x^0-iE_\p)(\partial_y^0+iE_\p)
\nonumber\\
&&\qquad\qquad\times\sum_{\lambda}
\epsilon_\lambda^\mu(\p)\epsilon_\lambda^\nu(\p)\;
\big<A_\mu(x)A_\nu(y)\big>\; .
\label{eq:AA}
\end{eqnarray}
Unsurprisingly, the operator $\big<A_\mu(x)A_\nu(y)\big>$ is identical
to what we considered previously in the single nucleus case.  In
particular, at leading order, the single gluon spectrum is evaluated
by replacing the two gauge operators in the right hand side of the
previous equation by classical solutions of the Yang-Mills
equations. These classical solutions are obtained by imposing retarded
boundary conditions that vanish in the remote past. The only
difference with the previous section and with eqs.~(\ref{eq:YM}) is
that the current $J^\nu$ that drives the solutions of the Yang-Mills
equations is now comprised of two contributions corresponding to each
of the nuclei. This is a significant complication in that, unlike the
single nucleus case, analytical solutions do not exist. However, the
classical fields and the inclusive spectrum have been computed
numerically~\cite{KrasnV1,KrasnV2,KrasnV3,KrasnNV1,KrasnNV2,KrasnNV3,Lappi1,Lappi2,Lappi3}.

Formally, the single inclusive gluon spectrum at leading order is a
functional of the LC gauge fields ${\cal A}_{1,2}$ of the two nuclei
on the surface $x^-=\epsilon$ and $x^+ = \epsilon$ respectively, or of
the covariant gauge fields $\wt{\cal A}^\pm_{1,2}$ in the strips
$0\leq x^- < \epsilon$ and $0\leq x^+ < \epsilon$ (see figure
\ref{fig:2nuc}),
\begin{equation}
\left.
E_\p\frac{dN}{d^3\p}
\right|_{_{\rm LO}}
\equiv{\cal O}_{_{\rm LO}}[{\cal A}_1,{\cal A}_2]
\equiv{\cal O}_{_{\rm LO}}[\wt{\cal A}^+_1,\wt{\cal A}^-_2]\; .
\label{eq:AA-LO}
\end{equation}
This quantity does not depend on the rapidity $y\sim\ln(p^+/p^-)$
because of the boost invariance of the classical equations of
motion~\cite{KovneMW1,KovneMW2,KovchR1}.

\subsection{One loop corrections}
At 1-loop, eq.~(\ref{eq:O-NLO}) can be used again to compute the
inclusive spectrum. The manipulations in sections \ref{sec:massageI}
and \ref{sec:massageII} were not specific to the case of a single
nucleus. Indeed, we did not specify the detailed content of the
current $J^\mu$ in section \ref{sec:nlo}. The only requirement for the
validity of the final formula is that one chooses an initial surface
$\Sigma$ which is locally space-like (or light-like at worst).

\begin{figure}[htbp]
\begin{center}
\resizebox*{!}{5cm}{\includegraphics{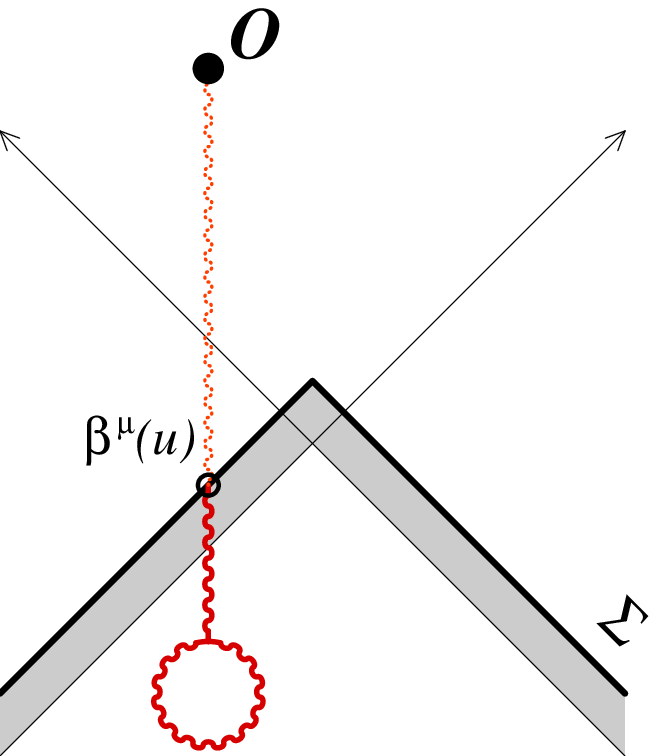}}
\hglue 10mm
\resizebox*{!}{5cm}{\includegraphics{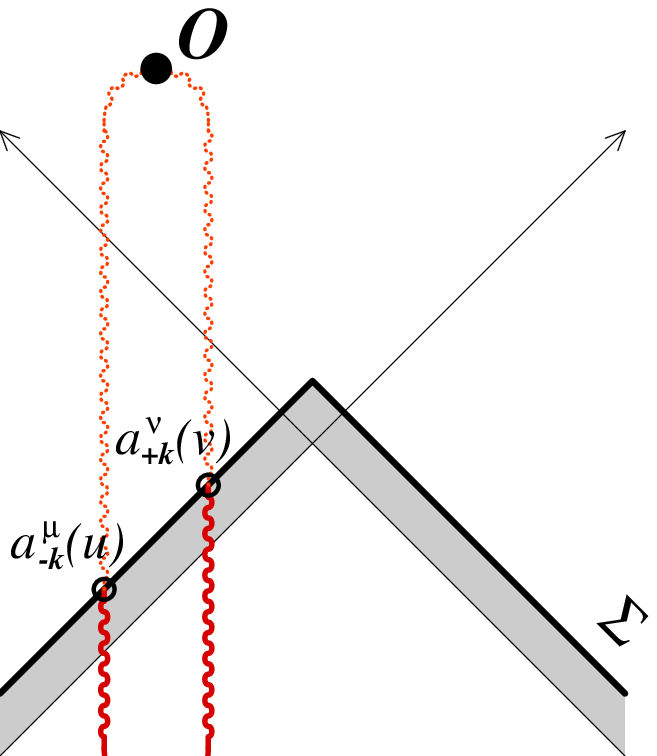}}
\end{center}
\caption{\label{fig:2nuc}NLO corrections in the collision of two
nuclei. The thick solid line is the initial surface where the
functions $\beta^\mu$ and $a_{\pm\k\lambda a}$ are evaluated. The
precise shape of the small portion of this surface located above the
forward light cone is not important because its contribution is
power suppressed.}
\end{figure}

We can now exploit this freedom in the choice of $\Sigma$ in order to
take a surface that treats the two nuclei on the same footing. A
convenient choice is a surface $\Sigma$ with the two branches
\begin{eqnarray}
&&
x^-=\epsilon\;, \quad x^+<\epsilon
\nonumber\\
&& 
x^+=\epsilon\;,\quad x^-<\epsilon\; , 
\end{eqnarray}
as illustrated by the thick solid line in figure \ref{fig:2nuc}.
We shall denote the measure on this initial surface as $d\Sigma_\u$.
It is simply $du^+ d^2\u_\perp$ on the first branch and $du^-
d^2\u_\perp$ on the second branch. Similarly, the definition of the
operator $[a\cdot\opt_\u]$ depends on the branch on which it is
evaluated, because the Green's formula for the classical fields
depends on a different set of initial field components on the two
branches\footnote{This result is evident from the derivation of the
  Green's formula in LC gauge discussed at length in appendix
  \ref{app:green}.}. It is also important to note that the functional
derivatives with respect to the initial gauge fields are derivatives
with respect to the field ${\cal A}_1$ of the first nucleus on the
first branch and likewise the field ${\cal A}_2$ of the second nucleus
on the second branch.

We need also to say a few words about the gauge in which the initial
fields on $\Sigma$ are expressed. On the left branch of $\Sigma$ (i.e.
on the branch $u^-=\epsilon$), we use the $A^+=0$ gauge, while we use
the $A^-=0$ gauge on the other branch. Using different gauge
conditions on these two branches is possible because they are not
causally connected. Similarly, for the propagation of the small
fluctuations $a_{\pm\k\lambda a}$ and $\beta$, we use the $A^+=0$
gauge if their endpoint is on the left branch of $\Sigma$, and the
$A^-=0$ gauge if it is on the other side.

Modulo these obvious changes, eq.~(\ref{eq:O-NLO1}) is valid in the
case of two nuclei and we can now express it as
\begin{eqnarray}
&&
{\cal O}_{_{\rm NLO}}=\Bigg[\,
\int\limits_{\Sigma}
d\Sigma_\u
\big[\beta\cdot\opt_\u\big]
\nonumber\\
&&\quad
+\frac{1}{2}\sum_{\lambda,a}
\int\!\!\frac{d^3\k}{(2\pi)^3 2E_\k}
\int\limits_{\Sigma}
d\Sigma_\u d\Sigma_\v
\big[a_{-\k \lambda a}\cdot\opt_\u\big]
\big[a_{+\k \lambda a}\cdot\opt_\v\big]
\Bigg]{\cal O}_{_{\rm LO}}[{\cal A}_1,{\cal A}_2]
\nonumber\\
&&\quad
+\Delta{\cal O}_{_{\rm NLO}}\; .
\label{eq:O-NLO2}
\end{eqnarray}
The first two terms in this formula are illustrated in figure
\ref{fig:2nuc}.  As in the case of a single nucleus, the
leading logs will cancel in $\Delta{\cal O}_{_{\rm NLO}}$ because of the charge conjugation 
symmetry discussed previously.

The leading log piece of the term involving $[\beta\cdot{\mathbbm
T}_\u]$ can be mapped into the corresponding term of the JIMWLK
equation in the same way as in the case of a single nucleus. Depending
on whether we are on the first or second branch of the initial surface
$\Sigma$, we get two terms which can be expressed together as
\begin{eqnarray}
&&
\Bigg[
\ln\left(\frac{\Lambda^+}{p^+}\right)
\int d^2\x_\perp\; \nu^b_1(\x_\perp)\;
\frac{\delta}{\delta \wt{\cal A}^+_{1,b}(\epsilon_{_Y},\x_\perp)}
\nonumber\\
&&+
\ln\left(\frac{\Lambda^-}{p^-}\right)
\int d^2\x_\perp\; \nu^b_2(\x_\perp)\;
\frac{\delta}{\delta \wt{\cal A}^-_{2,b}(\epsilon_{_Y},\x_\perp)}
\Bigg]\,{\cal O}_{_{\rm LO}}[\wt{\cal A}^+_1,\wt{\cal A}^-_2]\; ,
\label{eq:2nuc-nu}
\end{eqnarray}
where $\nu^b_{1,2}(\x_\perp)$ are respectively the one point functions
from the JIMWLK Hamiltonian for the two nuclei and likewise, $\wt{\cal
  A}^+_1,\wt{\cal A}^-_2$ are classical gauge fields in Lorenz gauge
of the first and second nucleus respectively. We have also introduced
a cutoff $\Lambda^-$, that separates the color sources of the second
nucleus from the dynamical fields.

There is a subtlety in generalizing the single nucleus derivation to
obtain this result. In eq.~(\ref{eq:up-int}), the integration over
$u^+$ runs from $-\infty$ to $+\infty$. Now, because of the choice of
the surface $\Sigma$, this integration runs only from $-\infty$ to
$0$, and we must justify that this difference is irrelevant.  To
simplify the notations in this argument, let us use the shorthand
$\alpha(u^+,\u_\perp)\equiv \partial_\mu (\Omega(u){\cal
  A}^\mu(u^+,\u_\perp))$. In our problem, the functional derivative
with respect to $\alpha(u^+,\u_\perp)$ is only applied to functionals
that depend solely on the $u^+$-independent mode of
$\alpha(u^+,\u_\perp)$, 
\begin{equation}
\alpha(\u_\perp)\equiv \frac{1}{L}\int du^+\;\alpha(u^+,\u_\perp)\; ,
\end{equation}
where $L$ is the length of the $u^+$ interval\footnote{Since here this
  interval is semi-infinite, it is best to consider $u^+\in[-L,0]$ in
  all the intermediate steps, and to take $L\to \infty$ only at the
  end.}. When this is the case, we have
\begin{equation}
\frac{\delta}{\delta\alpha(u^+,\u_\perp)}\;F[\alpha(\u_\perp)]
=\frac{1}{L}\;\frac{\delta}{\delta\alpha(\u_\perp)}\;F[\alpha(\u_\perp)]\; .
\end{equation}
Moreover, the result of this differentiation does not depend on the
value of $u^+$ in the l.h.s. Therefore, the subsequent integration
over $u^+$ merely generates a factor $L$ equal to the length of the
integration range. We have therefore proven that
\begin{equation}
\int du^+\;
\frac{\delta}{\delta\alpha(u^+,\u_\perp)}\;F[\alpha(\u_\perp)]
=\frac{\delta}{\delta\alpha(\u_\perp)}\;F[\alpha(\u_\perp)]\; ,
\end{equation}
regardless of the integration range for the variable $u^+$.

Another possible concern is whether there is a contribution to
$[\beta\cdot\opt_\u]$ from the small portion of the initial surface
$\Sigma$ that lies above the forward light cone in the region where
{\it both} $u^\pm$ are positive. It is easy to convince oneself that
the contribution from this region does not lead to stronger
singularities than the rest of the initial surface. Furthermore,
contributions from this region are phase space suppressed due to its
small size of order $\epsilon$.

The leading log contribution of the terms of eq.~(\ref{eq:O-NLO2})
that are bilinear in $[a\cdot\opt]$ is equally simple when the two
points $u$ and $v$ belong to the same branch of the initial surface
$\Sigma$. If this is so, it is straightforward to reproduce what we
did for a single nucleus, and we find the two separate contributions
\begin{eqnarray}
&&
\Bigg[
\ln\left(\frac{\Lambda^+}{p^+}\right)
\int d^2\x_\perp d^2\y_\perp\; \eta^{bc}_1(\x_\perp,\y_\perp)\;
\frac{\delta^2}{\delta \wt{\cal A}^+_{1,b}(\epsilon_{_Y},\x_\perp)
\delta \wt{\cal A}^+_{1,c}(\epsilon_{_Y},\y_\perp)
}
\nonumber\\
&&
\!\!\!\!\!\!\!\!\!\!
+
\ln\left(\frac{\Lambda^-}{p^-}\right)
\!\int \!
d^2\x_\perp d^2\y_\perp\, \eta^{bc}_2(\x_\perp,\y_\perp)\,
\frac{\delta^2}{\delta \wt{\cal A}^-_{2,b}(\epsilon_{_Y},\x_\perp)
\delta \wt{\cal A}^-_{2,c}(\epsilon_{_Y},\y_\perp)
}
\Bigg]{\cal O}_{_{\rm LO}}[\wt{\cal A}^\pm_{1,2}]\, .
\nonumber\\
&&
\label{eq:2nuc-eta}
\end{eqnarray}
Summing eqs.~(\ref{eq:2nuc-nu}) and (\ref{eq:2nuc-eta}), and
expressing $\nu$ in terms of $\eta$, we obtain the leading log 1-loop
expression for the single inclusive gluon spectrum to be
\begin{equation}
{\cal O}_{_{\rm NLO}}\empile{=}\over{\rm LLog}
\Bigg[
\ln\left(\frac{\Lambda^+}{p^+}\right)\,
{\cal H}_{1}
+
\ln\left(\frac{\Lambda^-}{p^-}\right)\,
{\cal H}_{2}
\Bigg]\,{\cal O}_{_{\rm LO}}[\wt{\cal A}^+_1,\wt{\cal A}^-_2]\; ,
\label{eq:O-NLO4}
\end{equation}
where ${\cal H}_{1,2}$ are the JIMWLK Hamiltonians of the first and
second nucleus respectively. This equation -- assuming we can prove
that there are no other terms at leading log -- is the generalization
of eq.~(\ref{eq:O-NLO3}) to the case of the collision of two nuclei.
In the next subsection, we will demonstrate that indeed there are no
other contributions.

\subsection{Absence of pre-collision mixings}
\begin{figure}[htbp]
\begin{center}
\resizebox*{!}{5cm}{\includegraphics{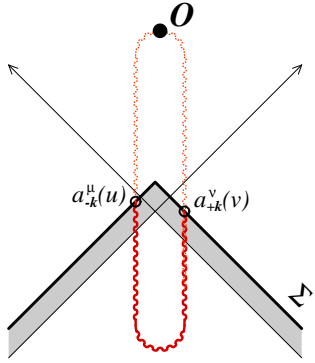}}
\end{center}
\caption{\label{fig:2nuc-1}Contribution that mixes the two nuclei and
may lead to a violation of factorization.}
\end{figure}
Thus far, we did not discuss the contribution to the bilinear
$[a\cdot\opt]$ terms where the coordinates $u$ and $v$ belong
to different branches of the initial surface. This contribution is
illustrated in figure \ref{fig:2nuc-1}. If it contains leading log
contributions, such a term would spoil eq.~(\ref{eq:O-NLO4}), because
it would generate a term that mixes derivatives with respect to
$\wt{\cal A}^+_1$ and $\wt{\cal A}^-_2$, thereby precluding any
possibility of factorization.

Fortunately, this possibility is not realized because terms where $u$
and $v$ are on different branches contain the phases
\begin{equation}
e^{ik^+(u^--v^-)}\,e^{ik^-(u^+-v^+)}\;
\end{equation}
in the integral over $d^3\k$. For generic points $u$ and $v$ in this
configuration, neither $u^--v^-$ nor $u^+-v^+$ are vanishing and these
exponentials oscillate rapidly when either $k^+\to +\infty$ or
$k^-\to+\infty$. Therefore, the integral over $k^+$ (or $k^-$) is
completely finite, and we do not get a large logarithm from this
configuration of $u$'s and $v$'s.

The only potential danger might come from the configuration where $u$
or $v$ (or both) lie in the small portion of $\Sigma$ above
the tip of the light cone. Again, such a configuration can at most
produce a logarithmic singularity, but is suppressed by a small phase
space prefactor of order $\epsilon$ due to the small size of this
region. Therefore, eq.~(\ref{eq:O-NLO4}) contains all the leading log
terms that show up in the 1-loop corrections to the single inclusive
gluon spectrum.

\subsection{Factorization}
Finally, integrating over all the configurations of the nuclear fields
$\wt{\cal A}^\pm_{1,2}$ with weights $W[\wt{\cal A}^+_1]$ and
$W[\wt{\cal A}^-_2]$, and using the fact that the JIMWLK Hamiltonian
is Hermitean, we can write the sum of the LO and NLO (leading logs
only) of the single inclusive gluon spectrum as
\begin{eqnarray}
&&
\left<{\cal O}_{_{\rm LO}}+{\cal O}_{_{\rm NLO}}\right>
\empile{=}\over{\rm LLog}
\int
\big[D\wt{\cal A}^+_1\big]\big[D\wt{\cal A}^-_2\big]\;
\Big\{
\Big[1+\Delta Y_1 {\cal H}_1\Big]\,
W\big[\wt{\cal A}^+_1\big]\Big\}\,
\nonumber\\
&&\qquad\qquad\qquad\qquad
\times
\Big\{
\Big[1+\Delta Y_2 {\cal H}_{2}\Big]\,
W\big[\wt{\cal A}^-_2\big]\Big\}
\,{\cal O}_{_{\rm LO}}[\wt{\cal A}^+_1,\wt{\cal A}^-_2]\; .
\end{eqnarray}
In this equation, we denote $\Delta Y_1\equiv\ln(\Lambda^+_1/p^+)$ and
$\Delta Y_2\equiv\ln(\Lambda_2^-/p^-)$, where $\Lambda^+$ is the
cutoff in the CGC description of the first nucleus, $\Lambda^-$ of the
second nucleus, and $p^\pm$ the longitudinal momentum components of
the produced gluon.  We can now choose the (arbitrary) cutoffs as
$\Lambda^\pm = p^\pm$ and express, as anticipated in
eq.~(\ref{eq:fact-formula}), the leading log part of the NLO result in
terms of the LO operator convoluted with the appropriately evolved
weight functions as
\begin{eqnarray}
&&
\left<{\cal O}\right>_{\rm LLog}
=
\int
\big[D\wt{\cal A}^+_1\big]\big[D\wt{\cal A}^-_2\big]\;
W_{_{Y_1}}\big[\wt{\cal A}^+_1\big]\,
W_{_{Y_2}}\big[\wt{\cal A}^-_2\big]\,
{\cal O}_{_{\rm LO}}[\wt{\cal A}^+_1,\wt{\cal A}^-_2]\; ,
\label{eq:fact-final}
\end{eqnarray}
where each of the $W[\wt{\cal A}^\pm]$'s obeys the JIMWLK equation
(possibly with different initial conditions if the two nuclei are not
identical) and 
$Y_1=\ln(P_1^+/p^+)$ and
$Y_2=\ln(P_2^-/p^-)$.

\section{High energy factorization result in context}
\label{sec:comp}
It is useful to consider our result in eq.~(\ref{eq:fact-final}) in
the context of related work in the high energy limit. Factorization,
in the specific sense of our work, was proven previously for
proton-nucleus collisions in the large $N_c$ limit of dipole
scattering off a large nucleus~\cite{KovchT1,JalilK1,JalilK2,KovneL1}.
In the case of nucleus-nucleus collisions, there has been recent work
by Braun, computing single and double inclusive gluon production in a
reggeon field theory approach~\cite{Braun1}. At present, it is unclear
how to relate these results to the JIMWLK evolution. A first attempt
at establishing such a dictionary between cut disconnected diagrams in
the CGC effective theory and cut Pomerons was discussed in
Ref.~\cite{GelisV1}; see also Refs.~\cite{KozloLP1,KozloLP2}.

It is important to note that the factorization theorem proven here is
valid only for inclusive quantities such as moments of the
multiplicity or energy distributions. In fact, it seems unlikely that
these results will extend to discussions of total cross-sections and
exclusive final states~\cite{Balit4,HattaIMST1,Hatta1}. Indeed, it is
known \cite{GelisV2,GelisV3} that the retarded nature of the boundary
conditions for the fields and field fluctuations has a close
connection with the inclusiveness of an observable, and we have seen
in the present paper that the retarded nature of these objects plays
an essential role in our proof of factorization. Whether the Pomeron
loops that may play a role in those computations are suppressed for
the observables we consider is also unclear. Our results certainly
suggest that these contributions are not important for inclusive
moments in nucleus-nucleus collisions, provided the densities
$\rho_{1,2}$ of color sources are large\footnote{If $\rho_{1,2}$ are
  not of order $g^{-1}$, then the power counting on which our
  considerations are based may be modified. Since it has been argued
  that Pomeron loops play a role in the dilute regime, this leaves
  open the possibility that these effects may alter our conclusions
  close to the fragmentation region of the projectiles.}.

Another important trend in the literature is computing next-to-leading
order contributions to high energy evolution. In the reggeon field
theory approach~\cite{Lipat2}, we note the very significant work on
multi-Regge factorization at NLO by Fadin and
collaborators~\cite{FadinKF1} which builds on the extension of the
BFKL equation to NLO~\cite{FadinL1,CamicC1,CiafaC1}. In the CGC
effective theory, there have been significant recent work to include
running coupling
corrections~\cite{Balit3,GardiKRW1,KovchW1,KovchW2,KovchW3,AlbacK1}
culminating in the recent NLO extension~\cite{BalitC1} of the
Balitsky-Kovchegov equation. As our result is valid for JIMWLK
factorization at leading log, these NLO results will be useful in
attempts to extend our proof of high energy factorization to
next-to-leading logarithmic accuracy.

Finally, we should emphasize that JIMWLK factorization proven here is
far more general and robust in comparison to the
$k_\perp$-factorization often discussed in the literature. The latter
is concerned with high energy factorization at the level of
unintegrated $k_\perp$ dependent parton
distributions~\cite{ColliE1,CatanCH1,LevinRSS1} and can be obtained in
the low density limit of JIMWLK
factorization~\cite{GelisV1,BlaizGV1,BlaizGV2}.
$k_\perp$-factorization also holds for single inclusive gluon
production at leading order in proton-nucleus
collisions~\cite{KovchT1,KovchM1,Braun2,ArmesB1,KopelST1,KharzKT1,BlaizGV1}.
$k_\perp$-factorization was however shown to be broken explicitly for
quark pair production even at leading order~\cite{BlaizGV2} albeit it
is restored~\cite{GelisV1} for large momenta $k_\perp \gg \qs$.
Likewise, this breaking of factorization is also seen for gluon pair
production~\cite{JalilK1,BaierKNW1}. Though JIMWLK factorization
remains to be proven for inclusive production of pairs, we anticipate
it is far more robust than $k_\perp$-factorization. 

To a large extent, factorization in hadronic collisions is merely a
consequence of causality~: two fast projectiles cannot interact
before they collide. Thus the objects that describe their
content must be universal --  independent of the other projectile,
and of the observable that one is going to measure {\sl after} the
collision.  However, this general argument does not tell us what
information should be included in the objects describing the
projectiles; indeed, this depends on the observable under
consideration, and on whether we are in the saturation regime or not.
In the saturated regime, a given observable will generally be produced
via the coherent interaction of many partons of the projectiles, which
means that one will need to know the probability of these multi-parton
configuration in the wavefunction of the projectiles. In contrast,
in the dilute regime, since only one parton of each projectile
interact, one needs only to know the probabilities for 1-parton
configurations. This is why JIMWLK factorization is more general than
$k_\perp$-factorization: the distribution $W[\rho]$ contain enough
information\footnote{It provides information about multiparton
  correlations such as $\big<\rho(x_1)\rho(x_2)\cdots\rho(x_n)\big>$.}
to calculate the non-integrated gluon distribution, but the converse
is certainly not true\footnote{Non integrated gluon distributions
  depend only on 2-parton correlations
  $\big<\rho(x_1)\rho(x_2)\big>$.}.  Similar considerations suggest
that JIMWLK factorization may not work in the case of exclusive
observables. Indeed, inclusive observable usually require less
detailed information about the projectiles than exclusive
ones\footnote{For instance, in order to study single diffractive
  processes, one would need ``conditional'' probabilities of
  multi-parton configurations, where one imposes the condition that no
  parton has been radiated between the rapidity of the projectile to
  the rapidity where the gap ends. This information is not provided by
  the distributions $W[\rho]$ that are the basis of JIMWLK
  factorization.}.

The factorization theorem that we have proved here is a necessary
first step before a full NLO computation of gluon production in the
Glasma. Eq.~(\ref{eq:fact-final}) includes only the NLO terms that are
enhanced by a large logarithm of $1/x_{1,2}$, while the complete NLO
calculation would also include the non enhanced terms. This would be
of the same order in $\as$ as the production of quark-antiquark
pairs \cite{GelisKL1,GelisKL2} from the classical field. Note that to
be really useful, this complete NLO calculation would probably have to
be promoted to a Next-to-Leading Log result by resumming all the terms
in $\as(\as\ln(1/x_{1,2})^n$. Now that evolution equations
in the dense regime are becoming available at NLO, work in this
direction is a promising prospect.

\section{Factorization, the Glasma and Thermalization}
\label{sec:glasma}
The Glasma is the non-equilibrium hot and dense matter formed
immediately in the aftermath of a high energy heavy ion
collision~\cite{LappiM1,KharzKV1,GelisV4}. How this matter thermalizes
is of great importance for a quantitative understanding of the
phenomenology of heavy ion collisions\footnote{Another important
  aspect is how jets propagate inside this matter, in order to assess
  issues such as leading parton quenching in jets.}.  We will discuss
here the relevance of our factorization theorem, present qualitative
ideas about its generalization and discuss their importance in
quantifying the properties of the Glasma.

At leading order, the Glasma is described by the solution of the
Yang-Mills equations in the forward light cone with retarded boundary
conditions (given by the classical fields of the two nuclei before the
collision). The produced fields have large occupation numbers of order
$\as^{-1}$ and are boost invariant~\cite{KovneMW1,KovneMW2}. This boost
invariance of fields implies that the classical dynamics can be
described by the proper time evolution of gauge fields that live in
the transverse plane. An interesting consequence of the classical
field dynamics is that the chromo-electric and magnetic fields are purely
longitudinal after the collision~\cite{KovneMW1,LappiM1} leading to
the generation of Chern-Simons charge density in the
collision~\cite{KharzKV1}. The Glasma fields at this order generate
only transverse pressure at proper times $\tau \gtrsim \qs^{-1}$ so it
seems impossible that a treatment of the Glasma at this order leads to
thermalization.

This is where the small quantum fluctuations of
the color field (of order $1$, compared to the classical
field of order $g^{-1}$) become relevant. In an observable such as the inclusive gluon
spectrum, these quantum fluctuations lead to corrections that are
$\as$ smaller than the leading order classical contribution.  As
we have discussed at length in the previous sections, some
contributions of these small fluctuations --- those that are enhanced by
leading powers of $\ln(1/x_{1,2})$ --- can be resummed and absorbed into
universal distributions $W[\rho]$ that describe the high energy
evolution of the nuclear wavefunctions.  

But what about the remaining part of these small fluctuation terms,
that are purely of order $\as$ relative to the classical fields?
Our resummation of leading logs corresponds to a well controlled
approximation provided the coefficients $d_{ni}$ in the expansion of
eqs.~(\ref{eq:expansion}) and (\ref{eq:dni}) are truly numbers of
order unity. Indeed, we have disregarded thus far the terms $d_{ni}$ for
$i<n$, on the basis that they do not have as many logs as powers of
$\as$. However, numerical simulations of the classical Yang-Mills
equations with initial conditions that break boost invariance
show the existence of an instability of the rapidity
dependent fluctuations~\cite{RomatV1,RomatV2,RomatV3}. In these
simulations, it is observed that the small rapidity dependent
perturbations superimposed to the boost invariant classical field grow
exponentially with the square root of time as \footnote{The fact that the
  square root of the proper time, rather than the proper time itself,
  controls the growth of the instability is due to the longitudinal
  expansion of the system. This has also been observed analytically in
  the study of the Weibel instability \cite{RebhaR1}.}
\begin{equation}
a^\mu\sim e^{\sqrt{\mu\tau}}\; ,
\end{equation}
where $\mu$ is a quantity of the order of $\qs$ (its precise value
depends on the wavelength of the fluctuation in the rapidity
direction). This growth has variously been interpreted as either a
Weibel type~\cite{RebhaR1,RomatV2} or Nielsen-Olesen
type~\cite{Iwaza1,FujiiI1} instability. The former mechanism in
particular has been discussed extensively as a possible mechanism for
thermalization in heavy ion
collisions~\cite{ArnolLM1,RebhaRS1,RebhaRS2,DumitNS1,MuellSW2,BodekR1,Mrowc1}.
The existence of these unstable modes suggests that our assumption
that the coefficients $d_{ni}$ for $i<n$ are of order unity is
incorrect.

Our present understanding is that there are three classes among the
small field fluctuations, that can be organized according to the 
momentum $p_\eta$ they have in the $\eta$ direction~:
\begin{itemize}
\item Zero modes ($p_\eta=0$) that generate a leading log. That the
  leading logs come solely from zero modes is obvious from the fact
  that the coefficients of the leading logs do not depend on $x^\pm$.
  These terms are already included in the resummation we have
  discussed at length in this paper.
\item Zero modes that do not contribute at leading log because they
  have an extra power of $k^-$ that prevents the divergence when
  $k^+\to\infty$ (see the discussion in section \ref{sec:log}). These
  terms have not been resummed in our scheme, and they do not seem to
  trigger the instability either. They would only become relevant in a
  full NLO calculation, and in resummation of Next-to-Leading Log
  terms~\cite{BalitC1}.
\item Non zero modes ($p_\eta\not=0$). These terms do not contribute
  large logarithms of $1/x_{1,2}$, but they are unstable and grow
  exponentially  as $\exp(\sqrt{\mu\tau})$.
\end{itemize}
It is the latter boost non-invariant terms that are potentially
dangerous. While also suppressed by a power of $\as$, they can be
enhanced by exponentials of the proper time after the collision.
Terms that diverge with time are called ``secular divergences'' and
some techniques for resumming these divergences are well
known\footnote{Indeed, one can think of the Boltzmann equation as an
  equation that effectively resums a certain class of secular
  divergences.} in the literature~\cite{Golde1}.  

Based on the above considerations, let us refine the expansion we
wrote in eqs~(\ref{eq:expansion}) and (\ref{eq:dni}), in order to keep
track also of powers of $\exp(\sqrt{\mu\tau})$. We should now write
\begin{equation}
{\cal O}\left[\rho_1,\rho_2\right]
=
\frac{1}{g^2}\Big[c_0+c_1 g^2+c_2 g^4+\cdots\Big]\; ,
\label{eq:expansion1}
\end{equation}
with
\begin{equation}
c_n\equiv\sum_{p=i}^n\sum_{i=0}^pf_{npi}\;e^{(p-i)\sqrt{\mu\tau}}\;\ln^i\left(\frac{1}{x_{1,2}}\right)\; .
\label{eq:cn2}
\end{equation}
In other words, the coefficients $d_{ni}$ that we have introduced in
eq.~(\ref{eq:dni}), and assumed to be of order unity, are in fact
\begin{equation}
d_{ni}=\sum_{p=i}^nf_{npi}\;e^{(p-i)\sqrt{\mu\tau}}\;,
\end{equation}
and can thus grow exponentially in time after the collision. In eq.~(\ref{eq:cn2}), the
sum of the number of logs and of factors $\exp(\sqrt{\mu\tau})$ (this
sum is the index $p$) cannot exceed $n$ at $n$ loops. This is because
a fluctuation mode cannot be at the same time a zero mode (required to
generate a log) and a non zero mode (required to generate an
instability). In this new language, the Leading Log resummation that
we have performed so far amounts to keep only the term $f_{nnn}$ in
every $c_n$.

At first sight, one may expect a complete breakdown of the Leading Log
description when the time
\begin{equation}
\tau_{\rm max}\sim \qs^{-1}\ln^2\left(\frac{1}{\as}\right)
\label{eq:taumax}
\end{equation}
is reached. This is the time at which 1-loop corrections become as
large as the LO contribution. This conclusion can be avoided if one
can resum these divergent contributions leading to a resummed result
that is better behaved for $\tau\to+\infty$.  Indeed, it is possible
to improve upon the Leading Log approximation, by keeping at every
loop order all the terms where $p=n$: this corresponds to all the
terms where every power of $\as$ is accompanied by either a log
or an $\exp(\sqrt{\mu\tau})$. Thus, let us define
\begin{equation}
{\cal O}_{_{\rm LLog+LInst}}\left[\rho_1,\rho_2\right]\equiv
\frac{1}{g^2}\sum_{n=0}^\infty g^{2n}\sum_{i=0}^n
f_{nni}\;e^{(n-i)\sqrt{\mu\tau}}\;\ln^i\left(\frac{1}{x_{1,2}}\right)\; .
\label{eq:improved-resum}
\end{equation}
The subscript ``LInst'' is meant for ``Leading Instability''.

In the formalism we have developed in this paper, the growth of small
fluctuations with time can be traced to the action of the linear
operator in eq.~(\ref{eq:O-NLO2}) on the classical field. The quantity
\begin{equation}
\opt_\u {\cal A}(x) \sim {\delta {\cal A}(x)\over \delta {\cal A}(\u)} 
\sim e^{\sqrt{\mu \tau}}
\; ,
\label{eq:diverg1}
\end{equation}
is a measure of how sensitive the classical field ${\cal A}(x)$ is to
initial condition at the point $\u$ on the initial surface. If there
is an instability, small perturbations of the initial conditions lead
to exponentially large deviations in the classical solutions. We will
assume for now that the improved resummation defined in
eq.~(\ref{eq:improved-resum}) can be performed and leads to
\begin{equation}
{\cal O}_{_{\rm LLog+LInst}}
=
Z[\opt_\u]\;{\cal O}_{_{\rm LLog}}[{\cal A}]\; ,
\end{equation}
where $Z[\opt_\u]$ is a certain functional of the operator $\opt_\u$.
In the r.h.s. we have emphasized the dependence of the observable on
the initial value of the gauge field. This formula can be expressed
more intuitively by performing a Laplace transform of $Z[\opt_\u]$
which reads
\begin{equation}
Z[\opt_\u]\equiv
\int\big[Da(\vec\u)\big]\;
e^{\int_\Sigma d^3\vec\u\;\big[a\cdot \opt_\u\big]}
\;
\widetilde{Z}[a(\vec\u)]\; .
\end{equation}
Given the structure of $a\cdot \opt_\u$ in eq.~(\ref{eq:T-def}), the
functional integration $[Da(\vec\u)]$ is an integration over the
initial fluctuation $a^\mu(\vec\u)$ itself and over some of its first
derivatives.  Because $\opt_\u$ is the generator of translations of
the initial conditions on the light cone, the exponential in the
previous formula is the translation operator itself. When this
exponential acts on a functional of the initial classical field ${\cal
  A}$, it gives the same functional evaluated with a shifted initial
condition ${\cal A}+a$. Therefore, we can write
\begin{equation}
{\cal O}_{_{\rm LLog+LInst}}
=\int\big[Da(\vec\u)\big]\;\widetilde{Z}[a(\vec\u)]\;
{\cal O}_{_{\rm LLog}}[{\cal A}+a]\; .
\end{equation}
The effect of the resummation is simply to add fluctuations to the
initial conditions of the classical field, with a distribution that
depends on the outcome of the resummation\footnote{In a recent work,
  using a completely different approach, the spectrum of initial
  fluctuations was found to be Gaussian\cite{FukusGM1}.}.  The
resummation lifts the limited applicability of the CGC approach
implied by eq.~(\ref{eq:taumax}). Indeed, after the resummation, the
fluctuation $a(\u)$ enters only in the initial condition for the full
Yang-Mills equations whose non-linearities prevent the solution from
blowing up. Combining our factorization formula in
eq.~(\ref{eq:fact-final}) with the {\it conjectured} result of the
resummation of the leading instabilities, one obtains a generalization
of eq.~(\ref{eq:fact-final}) which reads
\begin{eqnarray}
&&
\left<{\cal O}\right>_{\rm LLog+LInst}
=
\int
\big[D\wt{\cal A}^+_1\big]\big[D\wt{\cal A}^-_2\big]\;
W_{_{Y_1}}\big[\wt{\cal A}^+_1\big]\,
W_{_{Y_2}}\big[\wt{\cal A}^-_2\big]\,
\nonumber\\
&&\qquad\qquad\qquad\times
\int\big[Da(\vec\u)\big]\;\widetilde{Z}[a(\vec\u)]\;
{\cal O}_{_{\rm LO}}[\wt{\cal A}^+_1+a,\wt{\cal A}^-_2+a]
\; .
\label{eq:final}
\end{eqnarray}
This formula resums the most singular terms at each order in
$\as$. In comparison to the physics of the initial and final
state respectively in the collinear factorization framework, the
distributions $W[\rho]$ are analogous to parton distributions while
${\widetilde{Z}[a]}$ plays a role similar to that of a fragmentation
function\footnote{Naturally, this functional has nothing to do with a
  gluon fragmenting into a hadron. Instead, it describes how classical
  fields become gluons.}. To prove eq.~(\ref{eq:final}), and to
extract the spectrum of fluctuations, one needs to compute the
behavior of fluctuations on the forward light cone wedge at $x^\mp =
\epsilon, x^\pm \rightarrow +\infty$. 

Even after the resummations are performed in the initial and final
states, eq.~(\ref{eq:final}) still suffers from the usual problem of
collinear gluon splitting in the final state~\cite{KovchW3}. This
however is not a serious concern in heavy ion collisions because
collinear singularities occur only when one takes the $\tau\to+\infty$
limit. In practice, we expect to have switched to a more efficient 
description like kinetic theory or hydrodynamics long before this becomes a problem.
Indeed, the initial condition for hydrodynamics, which is specified in
terms of the energy-momentum tensor $T^{\mu\nu}$, is an infrared and
collinear safe quantity because it measures only the density and flow
of energy and momentum. It is straightforward to re-express our
results for multiplicity moments in terms of $T^{\mu\nu}$.

A far more challenging problem, that has still not received a
satisfactory answer, is to understand how the initial particle
spectrum -- or the local energy momentum-tensor -- become isotropic and perhaps even thermal.  Indeed, a very important question is
whether this improved resummation, that includes the leading unstable
terms, hastens the local thermalization of the system formed in heavy
ion collisions.

\section{Summary and outlook}
\label{sec:summary}
In this paper, we have presented a novel derivation of the JIMWLK
equation.  We showed that in this approach the JIMWLK Hamiltonian can
be determined entirely in terms of retarded propagators with no
ambiguities related to light cone pole prescriptions. Our approach
generalizes easily to the case of nucleus-nucleus collisions and we
were able to derive the factorization formula in
eq.~(\ref{eq:fact-final}). This formula is valid to all orders for
leading logs in $x$ and to all orders in the color charge densities of
the nuclei. For this factorization to work, it appears crucial to
consider an observable that can be expressed in terms of retarded
fields. Since we had previously linked retarded boundary conditions to
the inclusiveness of an observable, this emphasizes the importance of
inclusiveness for factorization, and the difficulties one may expect
when considering exclusive observables.

In view of this, it seems interesting to study whether the
factorization theorem proved here can be extended to less inclusive
quantities. One such example is the production of two jets that are
separated in rapidity by $\Delta Y \gg 1/\as$. In particular, can
the evolution between the jets be factorized from JIMWLK evolution of
the wavefunctions as in the case of inclusive gluon production?
Answers to these questions will be of great importance in assessing
whether the early time dynamics in heavy ion collisions leaves an
imprint in the long range rapidity correlations at later stages.

We further conjectured the existence of the generalized factorization
formula in eq.~(\ref{eq:final}). This expression also resums the
leading exponentials in time arising from the instability of the
classical fields to quantum fluctuations on the initial light cone
surface. The resulting spectrum of fluctuations is very important for
determining the subsequent thermalization of the Glasma. Work in this
direction is in progress.

\section*{Acknowledgements}
We would like to thank Nestor Armesto, Edmond Iancu, Jamal
Jalilian-Marian, Yuri Kovchegov, Alex Kovner, Misha Lublinsky and
Larry McLerran for very useful conversations. F. G and R. V thank the
Yukawa Institute for Theoretical Physics of Kyoto University and the
Yukawa International Program for Quark-Hadron Sciences for their
support during the completion of this work. R. V.'s work is supported
by the US Department of Energy under DOE Contract No.
DE-AC02-98CH10886. F.G.'s work is supported in part by Agence
Nationale de la Recherche via the programme ANR-06-BLAN-0285-01.

\appendix

\section{Gluon propagator in LC gauge}
\label{appL:Cprop}
\def\tn{\tilde{n}} Consider the QCD Lagrangian to which we add a gauge
fixing term proportional to $(\tn\cdot A)^2$,
\begin{equation}
{\cal L}\equiv-\frac{1}{4}F_{\mu\nu}^a F^{\mu\nu}_a
+\frac{1}{2\alpha}(\tn\cdot A)^2\; .
\end{equation}
We are mostly interested in the case where $\tn\cdot A=A^+$, but in
fact most of the discussion is valid for any vector $\tn^\mu$. In
order to determine the free propagator in this gauge, we need first to
isolate the quadratic part of the Lagrangian,
\begin{equation}
{\cal L}_{\rm quad}=\frac{1}{2}
A_\mu^a
\Big[\square g^{\mu\nu}-\partial^\mu\partial^\nu
+\frac{1}{\alpha}\tn^\mu\tn^\nu\Big]
A_\nu^a\; .
\label{eq:L-quad}
\end{equation}
The free propagator we are looking for is a Green's function of the
operator in the square brackets. Its calculation is best performed in
momentum space, where we need to invert
\begin{equation}
-g^{\mu\nu}k^2+k^\mu k^\nu+\frac{1}{\alpha}\tn^\mu\tn^\nu\; .
\label{eq:LC-kin}
\end{equation}
Because this tensor is symmetric in $(\mu,\nu)$, its inverse must be a
linear combination of $g^{\mu\nu}$, $k^\mu k^\nu$, $\tn^\mu\tn^\nu$
and $k^\mu \tn^\nu+k^\nu \tn^\mu$. Writing the most general general
linear combination of these elementary tensors, and multiplying it
with eq.~(\ref{eq:LC-kin}), we finally obtain the following expression
for the propagator in momentum space~:
\begin{equation}
D^{\mu\nu}_0(k)=
-\frac{g^{\mu\nu}}{k^2}
+\frac{k^\mu k^\nu}{(\tn\cdot k)^2}
\left[\alpha-\frac{\tn^2}{k^2}\right]
+\frac{k^\mu\tn^\nu+k^\nu \tn^\mu}{k^2(\tn\cdot k)}\; .
\end{equation}
Note that this expression is still incomplete, because we need to add
$i\epsilon$'s to the denominators in order to make the propagator
regular on the real energy axis. Doing so amounts to choosing certain
boundary conditions for the fields that evolve according to this
propagator. In this paper, the central object is the retarded
propagator, which has all its poles below the real energy axis. This
amounts to writing:
\begin{equation}
D^{\mu\nu}_{0,{}_R}(k)=
-\frac{g^{\mu\nu}}{k^2+ik^0\epsilon}
+\frac{k^\mu k^\nu}{(\tn\cdot k+i\epsilon)^2}
\left[\alpha-\frac{\tn^2}{k^2+ik^0\epsilon}\right]
+\frac{k^\mu\tn^\nu+k^\nu \tn^\mu}{(k^2+ik^0\epsilon)(\tn\cdot k+i\epsilon)}\; .
\label{eq:LC-prop-ret}
\end{equation}
(Our choice for the $i\epsilon$ prescription of the $\tn\cdot k$
denominators is indeed retarded if $n^0>0$. We will assume that this
is the case.)

In the case of the light-cone gauge $A^+=0$, this amounts to choosing
a vector $\tn^\mu$ that has $\tn^-=1$ and all its other components
zero. Moreover, we work in the ``strict'' light cone gauge, that
corresponds to the limit $\alpha\to 0$ for the gauge fixing parameter.
The propagator simplifies somewhat in this particular case~:
\begin{equation}
D^{\mu\nu}_{0,{}_R}(k)=
-\frac{1}{k^2+ik^0\epsilon}\left[g^{\mu\nu}
-
\frac{k^\mu\tn^\nu+k^\nu \tn^\mu}{\tn\cdot k+i\epsilon}\right]\; .
\label{eq:LC-prop-ret1}
\end{equation}
Note that this propagator is zero if any of its Lorentz indices is equal
to $+$.

\section{Green's formula in LC gauge}
\label{app:green}
An essential ingredient in our discussion is the Green's formula that
expresses a field fluctuation in terms of its value on some initial
surface. In this appendix, this initial surface will be the light-like
plane defined by $x^-=0$, but our derivation is more general than that
and applies to any initial surface.

\subsection{Green's formula for a small fluctuation in the vacuum}
Consider first a small field fluctuation $a^\mu$ propagating in the
vacuum. In the strict light cone gauge, it obeys
\begin{eqnarray}
&&
a^+(y)=0\; ,
\nonumber\\
&&
\big[\square_y g^{\mu\nu}-\partial_y^\mu\partial_y^\nu\big]a_\nu(y)=0\; .
\label{eq:fluct}
\end{eqnarray}
Recall also that the free propagator $D^{\rho\mu}_{0,{}_R}(x,y)$ obeys
\begin{equation}
D^{\rho}{}_{\mu\ 0,{}_R}(x,y)
\big[\stackrel{\leftarrow}{\square}_y g^{\mu\nu}
-\stackrel{\leftarrow}{\partial_y^\mu\partial_y^\nu}\big]
=g^{\rho\nu}\delta(x-y)\; ,
\label{eq:green-prop}
\end{equation}
where the arrows indicate that the derivatives act on the left.  Now,
multiply eq.~(\ref{eq:fluct}) by $D^{\rho\mu}_{0,{}_R}(x,y)$ on the
left, eq.~(\ref{eq:green-prop}) by $a_\nu(y)$ on the right, integrate
$y$ over all the domain defined by $y^->0$, and subtract the two
equations. One obtains
\begin{eqnarray}
a^\rho(x)=\int\limits_{y^->0} d^4y\;
D^{\rho}{}_{\mu\ 0,{}_R}(x,y)
\big[
\stackrel{\leftrightarrow}{\partial_y^\mu\partial_y^\nu}
-
\stackrel{\leftrightarrow}{\square}_y g^{\mu\nu}
\big]a_\nu(y)\; ,
\label{eq:green-tmp1}
\end{eqnarray}
where $\stackrel{\leftrightarrow}{A}\equiv
\stackrel{\rightarrow}{A}-\stackrel{\leftarrow}{A}$. Using the relations
\begin{eqnarray}
A\stackrel{\leftrightarrow}{\square}B
&=&
\partial^\mu\big[A\stackrel{\leftrightarrow}{\partial}_\mu B\big]\; ,
\nonumber\\
A \stackrel{\leftrightarrow}{\partial^\mu\partial^\nu} B
&=&
\frac{1}{2}
\partial^\mu
\big[
A \stackrel{\leftrightarrow}{\partial}{}^\nu B
\big]
+\frac{1}{2}
\partial^\nu
\big[
A \stackrel{\leftrightarrow}{\partial}{}^\mu B
\big]\; ,
\end{eqnarray}
we see that the integrand in eq.~(\ref{eq:green-tmp1}) is a total
derivative. Therefore, we can rewrite this integral as an integral on
the boundary of the integration domain. If the derivative we integrate
by parts is a $\partial^i$ or $\partial^-$, then the corresponding
boundary is located at infinity in the direction $y^i$ or $y^+$
respectively. We will assume that the field fluctuation under
consideration has a compact enough support so that these contributions
vanish. We are thus left with the terms coming from the derivative
$\partial^+$. The contribution from the boundary at $y^-=+\infty$ is
zero, because of our our choice of the retarded prescription for the
propagator. Therefore, the only contribution is from the boundary at
$y^-=0$,
\begin{equation}
a^\rho(x)=\!\!\int\limits_{y^-=0}\!\! dy^+d^2\y_\perp\;
D^{\rho}{}_{\mu\ 0,{}_R}(x,y)
\Big[
g^{\mu\nu}(n\cdot\stackrel{\leftrightarrow}{\partial}_y)
-
\frac{1}{2}\big(n^\mu
\stackrel{\leftrightarrow}{\partial}{}_y^\nu
+n^\nu
\stackrel{\leftrightarrow}{\partial}{}_y^\mu\big) 
\Big]a_\nu(y)\, ,
\label{eq:green-tmp2}
\end{equation}
where $n^\mu$ is a vector such that $n\cdot A=A^-$ (it is the unit
vector normal to the surface $y^-=0$). This formula indicates how the
value of the fluctuation at the point $x$ is related to its value on
an initial surface located at $y^-=0$ (Note that this dependence is
linear since small fluctuations obey a linear equation of motion). A
priori, it involves the values of all the components of the
fluctuation on this surface, as well as that of its first derivatives.
However, some of this information is not necessary because the
propagator vanishes when $\mu=+$ and because of the gauge condition
$a^+(y)=0$. If one eliminates from the previous formula all the terms
that are obviously zero and integrate some terms by parts\footnote{The
  antisymmetric derivatives
  $\stackrel{\leftrightarrow}{\partial}{}_y^-$ and
  $\stackrel{\leftrightarrow}{\partial}{}_y^i$ can be eliminated by
  integration by parts. This is not possible for
  $\stackrel{\leftrightarrow}{\partial}{}_y^+$ since the boundary term
  does not contain an integral with respect to $y^-$. This is why we
  have a term involving the derivative $\partial_y^+ D_{0,{}_R}^{\rho -}$.} ,
we get $a^\rho(x)\equiv {\cal B}_0^\rho[a](x)$, where ${\cal
  B}_0^\rho[a](x)$ is an integral that depends only on the value of
the field and of some of its derivatives on the initial surface,
\begin{eqnarray}
{\cal B}_0^\rho[a](x)
&=&
\int\limits_{y^-=0}dy^+d^2\y_\perp\;
\Big\{
\Big[\partial^y_\mu D_{0,{}_R}^{\rho \mu}(x,y)\Big]
a^-(y)
\nonumber\\
&&\qquad
-D_{0,{}_R}^{\rho -}(x,y)\Big[\partial_y^\mu a_\mu(y)\Big]
-D_{0,{}_R}^{\rho i}(x,y)\;2\partial_y^-\, a^i(y)
\Big\}\; .
\label{eq:green-tmp3}
\end{eqnarray}
Therefore, it appears that in the light-cone gauge $A^+=0$, and for an
initial surface $x^-=0$, we need to know the initial value of $a^-,
\partial^-a^i$ and $\partial_\mu a^\mu$ in order to fully determine
the value of the fluctuation at the point $x$. This fact is the reason
why there are only three terms in the definition of the operator
${\mathbbm T}_\u$ in eq.~(\ref{eq:T-def}) (but we postpone until the
end of this section the explanation of why one needs to include the
Wilson line $\Omega$ in this definition).

Moreover, the first term in the right hand side of
eq.~(\ref{eq:green-tmp3}) can be simplified considerably by using the
explicit expression of the free propagators in light cone gauge~:
\begin{eqnarray}
\partial^y_\mu D_{0,{}_R}^{\rho \mu}(x,y)
=
\delta^{\rho-}\,\theta(x^--y^-)\,\delta(x^+-y^+)\delta(\x_\perp-\y_\perp)\; .
\label{eq:div-prop}
\end{eqnarray}

\subsection{Green's formula for classical solutions}
There is also a similar Green's formula for retarded classical
solutions of the Yang-Mills equations. Contrary to the case of small
fluctuations, we do not assume that the gauge field is small, and we
keep all the self-interactions as well as the interactions with some
external source. Formally, we can write the Lagrangian as
\begin{equation}
{\cal L}={\cal L}_{\rm quad}-U({\cal A})\; ,
\end{equation}
where $U({\cal A})$ is a local polynomial of the gauge field. It contains the
3- and 4-gluon couplings and the coupling to the external source. In
the ${\cal A}^+=0$ gauge, the corresponding classical equation of motion is 
\begin{equation}
\big[\square_y g^{\mu\nu}-\partial_y^\mu\partial_y^\nu\big]{\cal A}_\nu(y)=\frac{\partial U({\cal A})}{\partial {\cal A}_\mu(y)}\; .
\end{equation}
Then one can follow the same procedure as in the case of small
fluctuations, and we obtain
\begin{eqnarray}
{\cal A}^\rho(x)=
\int\limits_{y^->0} d^4y\;
D^{\rho\mu}_{0,{}_R}(x,y)\,
\frac{\partial U({\cal A})}{\partial {\cal A}^\mu(y)}
+{\cal B}^\rho_0[{\cal A}](x)\; .
\label{eq:green-class}
\end{eqnarray}
Of course, the dependence of the classical field on its initial
conditions is no longer linear because of the first term in the right
hand side; the self interactions of the gauge fields lead to an
involved dependence on the initial conditions.

\subsection{Green's formula for $a^\mu$ in a background field}
Finally, the Green's formula of eq.~(\ref{eq:green-tmp3}) can be
extended to the situation where the fluctuation $a^\mu(x)$ propagates
on top of a classical background field ${\cal A}^\mu$ rather than the
vacuum. The only change is that the free propagator must be replaced
by the propagator in a background field. The property that its $\mu=+$
Lorentz component vanishes remains true, because it is a consequence
of the choice of the gauge. For such a fluctuation, there is also a
Green's formula that uses only the free gauge propagator, and where
the interactions with the background field appear explicitly as the
additional term
\begin{eqnarray}
a^\rho(x)=
\int\limits_{y^->0}d^4y\;D^{\rho\nu}_{0,{}_R}(x,y)
\frac{\partial^2 U({\cal A})}{\partial {\cal A}_\nu(y)\partial A^\sigma(y)}
\;
a^\sigma(y)+{\cal B}^\rho_0[a](x)\; .
\label{eq:green-tmp4}
\end{eqnarray}
The derivation of this formula is very similar to that for the
classical field ${\cal A}^\mu$. We can also rewrite it in a form very
similar to eq.~(\ref{eq:green-tmp3}), i.e. $a^\rho(x)={\cal B}[a](x)$
with
\begin{eqnarray}
{\cal B}^\rho[a](x)
&=&
\int\limits_{y^-=0}dy^+d^2\y_\perp\;
\Big\{
\Big[\partial^y_\mu D^{\rho \mu}_{_R}(x,y)\Big]
a^-(y)
\nonumber\\
&&\qquad\quad
-D^{\rho -}_{_R}(x,y)\Big[\partial_y^\mu a_\mu(y)\Big]
-D^{\rho i}_{_R}(x,y)\;2\partial_y^-\, a^i(y)
\Big\}\; .
\label{eq:green-tmp5}
\end{eqnarray}
The boundary term ${\cal B}[a]$ differs from ${\cal B}_0[a]$ in the
fact that it contains the retarded propagator $D^{\mu\nu}_{_R}$ {\sl
  dressed by the background field } instead of the bare retarded
propagator $D^{\mu\nu}_{0,{}_R}$. A crucial difference between the
dressed and bare propagators is that the simplification of
eq.~(\ref{eq:div-prop}) does not occur with the dressed propagator.

In the derivation of the JIMWLK equation, the fluctuations $a^\mu(x)$
one considers are fluctuations whose initial condition at
$x^0\to-\infty$ are plane waves of momentum $k$. One can calculate
explicitly their value on the initial surface, which means that we
know analytically the quantities $a^-$, $\partial^- a^i$ and
$\partial_\mu a^\mu$ in the r.h.s. of eq.~(\ref{eq:green-tmp5}).  A
crucial property is that the initial values of $a^-$ and $\partial^-
a^i$ are suppressed by an extra factor $1/k^+$, and thus any term
containing them cannot have a logarithmic divergence when
$k^+\to+\infty$. This argument is correct provided the prefactors of
these quantities in eq.~(\ref{eq:green-tmp5}) do not bring factors of
$k^+$. There is no problem with the second and third terms, since
their prefactors is just a propagator.

However, as we shall see now, the coefficient of the first term can be
large because it involves the derivative of the propagator. The only
case of practical interest to us is when the background field above
the initial surface is a pure gauge field such as the one given in
eq.~(\ref{eq:class-LC}). In this particular case, there is a simple
relationship between the dressed and bare propagators~:
\begin{equation}
D^{\rho\mu}_{_R}(x,y)=\Omega^\dagger(x)\;D^{\rho\mu}_{0,{}_R}(x,y)\;\Omega(y)\; .
\end{equation}
This can be seen by applying a gauge transformation $\Omega^\dagger$
to the problem, which has the effect of removing the pure gauge
background. Using this equation, as well as eq.~(\ref{eq:div-prop}),
we now obtain
\begin{equation}
\partial_\mu^y D^{\rho\mu}_{_R}(x,y)
=
\Omega^\dagger(x)\;\Big[\partial_\mu^y D^{\rho\mu}_{0,{}_R}(x,y)\Big]\;\Omega(y)
+
D^{\rho\mu}_{_R}(x,y)\;\Omega^\dagger(y)\partial_\mu^y \Omega(y)
\end{equation}
The problem is that we take the derivative of the Wilson line
$\Omega(y)$ in a region where it is changing very quickly. Only the
term with the $\partial_y^+$ derivative exhibits this issue (since the
large derivatives are those in the $y^-$ direction),
\begin{equation}
D^{\rho-}_{_R}(x,y)\;\Omega^\dagger(y)\partial^+_y\Omega(y)\; .
\end{equation}
{}From its structure, it is obvious that this term mixes with the second
term in the r.h.s. of eq.~(\ref{eq:green-tmp5}) (which, as explained
in section \ref{sec:log}, leads to a logarithmic divergence); it would
thus be incorrect to keep the latter while not considering the former.
There are two ways to deal with this issue: keep track separately of
these two terms, or try to combine them into a single term.  The
second option is the simplest, and from the above considerations, we
know how to achieve it: by rotating the fluctuation $a^\mu$, $a^\mu\to
\Omega a^\mu$, we can rewrite the boundary term as
\begin{eqnarray}
&&{\cal B}^\rho[a](x)
=
\Omega^\dagger(x)
\int\limits_{y^-=0}dy^+d^2\y_\perp\;
\Big\{
\Big[\partial^y_\mu\big(D^{\rho \mu}_{0,{}_R}(x,y)\Big]
\Omega(y)a^-(y)
\nonumber\\
&&\qquad
-D^{\rho -}_{0,{}_R}(x,y)\Big[\partial_y^\mu \Omega(y)a_\mu(y)\Big]
-D^{\rho i}_{0,{}_R}(x,y)\;2\partial_y^-\, \Omega(y)a^i(y)
\Big\}\; ,
\label{eq:green-tmp6}
\end{eqnarray}
where we have now only bare propagators. This is why the most
convenient definition of ${\mathbbm T}_\u$ in eq.~(\ref{eq:T-def})
involves functional derivatives with respect to $\Omega a^\mu$ rather
than $a^\mu$ itself\footnote{Of course, the two ways of defining
  ${\mathbbm T}_\u$ --with and without the $\Omega$-- are exactly
  equivalent.  But if we did not include the $\Omega$ in the
  definition, the logarithmic divergences would come from a
  combination of the second and third terms of eq.~(\ref{eq:T-def}),
  instead of being limited to the third term if we include the
  $\Omega$ in the definition of ${\mathbbm T}_\u$.}.  Note that for
this discussion to hold, it is only necessary that the background
field is a pure gauge in the vicinity above the initial surface, since
the derivative is with respect to a coordinate on this initial
surface.  Whether the background field is a pure gauge everywhere
above the initial surface is not important.

\section{Two-dimensional free propagator}
\label{app:2d-prop}
In the derivation of the JIMWLK equation, one makes use of several
formulas involving the bare two-di\-men\-sio\-nal propagators. These
formulas are not new~: all of them have already been used in one form
or another in previous papers discussing the JIMWLK equation. We
compile them in this appendix, with their derivation, as a convenient
reference for the reader.

Let us denote $G(\x_\perp-\y_\perp)$ a Green's function of the
2-dimensional Laplacian operator,
\begin{equation}
{\bs\partial}_\perp^2\,G(\x_\perp-\y_\perp)=\delta(\x_\perp-\y_\perp)\; .
\end{equation}
It admits a simple Fourier representation,
\begin{equation}
G(\x_\perp-\y_\perp)
=
-\int\frac{d^2\k_\perp}{(2\pi)^2}\;
e^{i\k_\perp\cdot(\x_\perp-\y_\perp)}\;\frac{1}{\k_\perp^2}\; .
\end{equation}
Note that this object suffers from an infrared problem, which is
obvious for dimensional reasons: this propagator is a dimensionless
object in coordinate space, invariant under translations and
rotations, and therefore it must be a function of
$\mu\big|\x_\perp-\y_\perp\big|$ where $\mu$ is some mass scale that
was not present in the previous equation.

Derivatives of this propagator do not suffer from this infrared
ambiguity. Consider for instance\footnote{Let us recall that
  $\partial_x^i=\frac{\partial}{\partial
    x_i}=-\frac{\partial}{\partial x^i}$.}
\begin{equation}
\partial_x^i \, G(\x_\perp-\y_\perp)
=
i\int\frac{d^2\k_\perp}{(2\pi)^2}\;
e^{i\k_\perp\cdot(\x_\perp-\y_\perp)}\;\frac{k^i}{\k_\perp^2}\; .
\end{equation}
{}{}From its symmetries and dimension, it is obvious that this derivative
can be written as
\begin{equation}
\partial_x^i \, G(\x_\perp-\y_\perp)
=
C\;\frac{\x_\perp^i-\y_\perp^i}{(\x_\perp-\y_\perp)^2}\; ,
\end{equation}
where the prefactor $C$ is dimensionless. Because the derivative of
the propagator is not infrared singular, the cutoff $\mu$ cannot
appear in its expression and $C$ must be a pure number (otherwise it
would have to be a function of $\mu\big|\x_\perp-\y_\perp\big|$ to
have the correct dimension). In order to determine the constant, take
another derivative $\partial_x^i$ and integrate over $\x_\perp$ the
resulting equation over some domain $\Omega$ of the plane that
contains the point $\y_\perp$. On the left hand side, we get the
integral of a delta function since $G$ is a Green's function of
${\bs\partial}_\perp^2$.  We then get
\begin{equation}
1=C\;\int_\Omega d^2\x_\perp\;
\partial_x^i\,\frac{\x_\perp^i-\y_\perp^i}{(\x_\perp-\y_\perp)^2}\; .
\end{equation}
The right hand side can be transformed by using the 2-dimensional
Stokes theorem, leading to an integral on the boundary of $\Omega$
(oriented counter-clockwise)
\begin{equation}
  1=C\;\int_{\partial\Omega}
\frac{\epsilon^{ij}\,(\x_\perp^i-\y_\perp^i)\,dx^j}{(\x_\perp-\y_\perp)^2}\; ,
\end{equation}
where $\epsilon^{ij}$ is completely antisymmetric ($\epsilon^{12}=1$).
The contour integral in this equation is a topological quantity, that
depends only on the winding number of the contour $\partial\Omega$
around the point $\y_\perp$. Thus, it is best calculated by deforming
$\partial\Omega$ into the unit circle around the point $\y_\perp$. We
get easily
\begin{equation}
1=2\pi C\; .
\end{equation}
Thus we have
\begin{equation}
\partial_x^i \, G(\x_\perp-\y_\perp)
=
\frac{1}{2\pi}\;\frac{\x_\perp^i-\y_\perp^i}{(\x_\perp-\y_\perp)^2}\; .
\label{eq:deriv-2Dgreen}
\end{equation}

The second derivative of the propagator is also useful in the
derivation of the JIMWLK equation. By applying $\partial_x^j$ to the
previous equation, one obtains
\begin{eqnarray}
\partial_x^i\partial_x^j\,G(\x_\perp-\y_\perp)
&=&
\frac{1}{2\pi}\;\partial_x^j\,
\frac{\x_\perp^i-\y_\perp^i}{(\x_\perp-\y_\perp)^2}
\nonumber\\
&=&
\frac{1}{2\pi(\x_\perp-\y_\perp)^2}
\;
\left[
\delta^{ij}
-
2\frac{(\x_\perp^i-\y_\perp^i)(\x_\perp^j-\y_\perp^j)}{(\x_\perp-\y_\perp)^2}
\right]\; .
\end{eqnarray}
This formula, although perfectly correct for $\x_\perp\not=\y_\perp$,
is incorrect at the point $\x_\perp=\y_\perp$. In order to see this,
take the trace over the indices $i$ and $j$. In the left hand side, we
have the Laplacian of the propagator, i.e.
$\delta(\x_\perp-\y_\perp)$, while the right hand side would give
zero.  Thus the full formula for the second derivative is 
\begin{equation}
\partial_x^i\partial_x^j\,G(\x_\perp\!-\!\y_\perp)
=
\frac{\delta^{ij}}{2}\delta(\x_\perp\!-\!\y_\perp)
+
\frac{1}{2\pi}\Delta^{ij}(\x_\perp-\y_\perp)\; ,
\label{eq:dijG}
\end{equation}
with
\begin{equation}
\Delta^{ij}(\x_\perp-\y_\perp)
\equiv
\frac{1}{(\x_\perp\!-\!\y_\perp)^2}
\;
\left[
\delta^{ij}
\!-\!
2\frac{(\x_\perp^i\!-\!\y_\perp^i)(\x_\perp^j\!-\!\y_\perp^j)}{(\x_\perp\!-\!\y_\perp)^2}\right]\; .
\label{eq:dijG1}
\end{equation}
This function $\Delta^{ij}$ obeys an interesting identity. By
integration by parts, one can check that
\begin{eqnarray}
&&
\int \frac{d^2\u_\perp}{(2\pi)^2}\frac{d^2\v_\perp}{(2\pi)^2}\;
\frac{(\x_\perp^i-\u_\perp^i)(\y_\perp^j-\v_\perp^j)}{(\x_\perp-\u_\perp)^2(\y_\perp-\v_\perp)^2}\;
\partial_u^i\partial_v^j G(\u_\perp-\v_\perp)
=
\nonumber\\
&&\qquad
=
-\frac{1}{(2\pi)^2}\int \frac{d^2\u_\perp}{(2\pi)^2}\;
\frac{(\x_\perp^i-\u_\perp^i)(\y_\perp^i-\u_\perp^i)}{(\x_\perp-\u_\perp)^2(\y_\perp-\u_\perp)^2}
\nonumber\\
&&\qquad
=
-\int \frac{d^2\u_\perp}{(2\pi)^2}\frac{d^2\v_\perp}{(2\pi)^2}\;
\frac{(\x_\perp^i-\u_\perp^i)(\y_\perp^j-\v_\perp^j)}{(\x_\perp-\u_\perp)^2(\y_\perp-\v_\perp)^2}\;
\delta^{ij}\delta(\u_\perp-\v_\perp)\; .
\end{eqnarray}
Using now eq.~(\ref{eq:dijG}), we obtain the following identity,
\begin{equation}
\int \frac{d^2\u_\perp}{(2\pi)^2}\frac{d^2\v_\perp}{(2\pi)^2}\;
\frac{(\x_\perp^i-\u_\perp^i)(\y_\perp^j-\v_\perp^j)}{(\x_\perp-\u_\perp)^2(\y_\perp-\v_\perp)^2}\,\Big[
\frac{\delta^{ij}}{2}\delta(\u_\perp-\v_\perp)
-
\frac{1}{2\pi}\Delta^{ij}(\u_\perp-\v_\perp)
\Big]=0\; .
\end{equation}

Let us also provide an alternate representation of the 2-dimensional
propagator that is sometimes helpful. Let us start with the integral
\begin{equation}
\int \frac{d^2\u_\perp}{(2\pi)^2}\,
\frac{\u_\perp^i-\x_\perp^i}{(\u_\perp-\x_\perp)^2}
\frac{\u_\perp^i-\y_\perp^i}{(\u_\perp-\y_\perp)^2}
=
\int d^2\u_\perp\,
\Big[\partial_u^i G(\u_\perp-\x_\perp)\Big]
\Big[\partial_u^i G(\u_\perp-\y_\perp)\Big]\; .
\end{equation}
The integral in the right hand side can be performed by parts, since
it leads to the Laplacian of a propagator, which is a delta function.
Thus, we obtain the identity
\begin{equation}
G(\x_\perp-\y_\perp)
=
-
\int \frac{d^2\u_\perp}{(2\pi)^2}\,
\frac{\u_\perp^i-\x_\perp^i}{(\u_\perp-\x_\perp)^2}
\frac{\u_\perp^i-\y_\perp^i}{(\u_\perp-\y_\perp)^2}
\; .
\label{eq:G-alt}
\end{equation}
Note that the integral over $\u_\perp$ suffers from the same infrared
problems that we have already mentioned at the beginning of this
appendix.


\end{document}